\documentclass[final,5p,times,twocolumn]{elsarticle}
\usepackage{amssymb}
\usepackage{graphicx}
\usepackage[space]{grffile}
\usepackage{latexsym}
\usepackage{textcomp}
\usepackage{longtable}
\usepackage{multirow,booktabs}
\usepackage{amsfonts,amsmath,amssymb}
\usepackage{natbib}
\usepackage{url}
\usepackage{hyperref}
\hypersetup{colorlinks=false,pdfborder={0 0 0}}
\usepackage[english]{babel}

\newcommand{\makefigure}[2]{
\begin{figure}[ht!]
\begin{center}
\includegraphics[width=0.9\columnwidth]{#1.eps}
\caption{#2 \label{#1}}
\end{center}
\end{figure}}

\newcommand{\dreport}{DojoReport}
\newcommand{\ourtable}{PD-PBE}
\newcommand{\df}{$\Delta$-Gauge}
\newcommand{\dfp}{$\Delta$'-Gauge}

\journal{Computer Physics Communications}

\begin{document}

\begin{frontmatter}

\begin{keyword}
First-principles  calculation \sep Electronic structure
\sep Density Functional Theory \sep Pseudopotential
\end{keyword}

\title{The \textsc{PseudoDojo}: Training and grading a 85 element optimized norm-conserving pseudopotential table}

\address[ucl]{Nanoscopic Physics, Institute of Condensed Matter and Nanosciences, Universit\'{e} Catholique de Louvain, 1348 Louvain-la-Neuve, Belgium}
\address[etsf]{European Theoretical Spectroscopy Facility (ETSF)}
\address[ulg]{Q-Mat, Department of Physics, University of Li\`{e}ge (Belgium)}
\address[ru]{Department of Physics and Astronomy, Rutgers University, Piscataway, NJ 08854-8019, USA}
\address[msr]{Mat-Sim Research LLC, P. O. Box 742, Murray Hill, NJ, 07974, USA}

\author[ucl,etsf]{M. J. {van Setten}}
\author[ucl,etsf]{M. Giantomassi}
\author[ulg,etsf]{E. Bousquet}
\author[ulg,etsf]{M. J. Verstraete}
\author[ru,msr]{D. R. Hamann}
\author[ucl,etsf]{X. Gonze}
\author[ucl,etsf]{G.-M. Rignanese}

\begin{abstract}
First-principles calculations in crystalline structures are often performed with a planewave basis set. To make the number of basis functions tractable two approximations are usually introduced: core electrons are frozen and the diverging Coulomb potential near the nucleus is replaced by a smoother expression.
The norm-conserving pseudopotential was the first successful method to apply these approximations in a fully ab initio way.  Later on, more efficient and more exact approaches were developed based on the ultrasoft and the projector augmented wave formalisms. These formalisms are however more complex and developing new features in these frameworks is usually more difficult than in the norm-conserving framework. 
Most of the existing tables of norm-conserving pseudopotentials, generated long ago, do not include the latest developments, are not systematically tested or are not designed primarily for high accuracy.
In this paper, we present our \textsc{PseudoDojo} framework for developing and testing full tables of pseudopotentials, and demonstrate it with a new table generated with the \textsc{ONCVPSP} approach.
The \textsc{PseudoDojo} is an open source project, building on the \textsc{AbiPy} package, for developing and systematically testing pseudopotentials. At present it contains 7 different batteries of tests executed with \textsc{ABINIT}, which are performed as a function of the energy cutoff. The results of these tests are then used to provide hints for the energy cutoff for actual production calculations. 
Our final set contains 141 pseudopotentials split into a standard and a stringent accuracy table. In total around 70.000 calculations were performed to test the pseudopotentials. 
The process of developing the final table led to new insights into the effects of both the core-valence partitioning and the non-linear core corrections on the stability, convergence, and transferability of norm-conserving pseudopotentials.   
The \textsc{PseudoDojo} hence provides a set of pseudopotentials and general purpose tools for further testing and development, focusing of highly accurate calculations and their use in the development of ab initio packages. The pseudopotential files are available on the \textsc{PseudoDojo} web-interface www.pseudo-dojo.org in the psp8, UPF2, and PSML 1.1 formats.

\end{abstract}

\end{frontmatter}

\bibliographystyle{elsarticle-num}

\section{Introduction}\label{sect:introduction}
Many physical and chemical properties of solids are determined by the structure and dynamics of the valence electrons. This is true in particular for the formation of chemical bonds, but also for the magnetic behavior and for low-energy excitations. In contrast, the core electrons only indirectly affect these properties. Based on these observations, Density Functional Theory (DFT) electronic structure calculations often assume that the complicated interaction between valence electrons and the ions (formed by the atom nuclei and the core electrons) can be replaced by an effective potential known as a pseudopotential (PSP). The core states are thus eliminated and the valence electrons are described by smooth pseudo-wavefunctions. This is particularly useful when a planewave (PW) basis set is used to describe the electronic wavefunctions. Such a basis set has the nice advantage that its completeness can be systematically improved thanks to a unique parameter, the maximal kinetic energy of the planewaves in the basis set, also called the energy cut-off ($E_\mathrm{c}$). Describing the oscillations of the all-electron (AE) wavefunctions near the atomic core would indeed require a prohibitively large number of planewaves. 

One can safely state that any calculation using pseudopotentials can only be as efficient and accurate as the pseudopotentials that are used. Obviously, the problem of finding good pseudopotentials could be avoided altogether by using a basis set that is capable of describing all electronic states on an equal footing. The all-electron approaches, however, immediately lose the elegance of the single convergence parameter in the planewave approach. In a sense the problem of finding a good pseudopotential is now moved to finding a good basis set. Recently it was shown that indeed the variations between the results obtained with different AE-codes can be as large as the differences between the results of AE-codes and PW-codes.~\cite{Lejaeghere2016}

Norm-conserving pseudopotentials (NCPPs)~\cite{Hamann_1979, Bachelet_1982} are among the first pseudopotentials that were routinely used in realistic calculations and paved the way for the ever expanding application of density functional theory~\cite{PhysRev.136.B864, PhysRev.140.A1133} to solids. It is because of the elegance of the norm-conserving approach that NCPPs are supported by many {\em ab-initio} codes. The relatively simple and robust formalism of the NCPP also means that new developments are usually implemented for NCPPs first, see e.g. the recent availability of the temperature dependence of the electronic structure~\cite{Ponce2014}.

Unfortunately, many NCPP tables still in use nowadays for first-principles calculations were generated long ago, before the advent of optimization techniques such as the RRKJ method.~\cite{Rappe_1990} Even more importantly, no systematic validation of these tables is available. Very few of these pseudopotentials allow one to perform non-collinear calculations with the inclusion of the spin-orbit (SO) term. Last but not least, most of these legacy NCPPs employ only one projector per angular channel, hence it is difficult to find NCPPs including semi-core states or pseudopotentials with good scattering properties at high energies.

Presently many NPCC tables are available: The HGHK tables~\cite{Hartwigsen98,Krack2005} provides spin orbit coupling but it was not primarily designed for PW applications and, indeed, on average rather large cutoffs are needed. The NC tables previously available for use with \textsc{ABINIT},\footnote{Originally designed by Allan and Klein and later extended by one of us (MJV) for reference still provided on the \textsc{ABINIT} web-site under 'previous atomic datasets'} contain semi-core states for selected elements but the input files and the pseudopotential generator are not available anymore. The tables from the OPIUM project~\cite{OPIUM} have RRKJ optimization but not all the atoms of the periodic table  are available and multiple projectors for a given angular channel are not supported. NC potentials from the QUANTUM-ESPRESSO community are available~\cite{QE-PP} in the UPF format, but do not have more than one projector. The SG15 table~\cite{Schlipf2015} was designed for efficiency and does not have non-linear core corrections.

As compared to the more recently developed ultrasoft pseudopotentials~\cite{Vanderbilt_1990} (USPP) and the projector-augmented-wave method~\cite{Bl_chl_1994} (PAW), calculations using NCPPs usually require a larger kinetic energy cutoff making them less efficient. The implementation of both the PAW and USPP formalisms is however much more demanding. Moreover little is known about the reliability of these two approaches when applied beyond standard ground state calculations.~\cite{PhysRevB.90.075125} In contrast NCPPs have been used for decades in different ab-initio fields.  

NCPPs continue to represent a valid choice for ab-initio calculations because of the simplicity and robustness of the formalism. We also believe that many future developments in first principles codes will be first implemented within the NCPP formalism and eventually generalized to the USPP/PAW case (NCPPs can be seen as a particular case of the USPP/PAW formalism under certain assumptions).
For all the reasons mentioned above the ab-initio community would greatly benefit from the availability of a periodic table of reliable and accurate NCPPs. 

With this in mind, we have constructed a new NCPP table, using the PBE exchange-correlation functional\cite{Perdew1996}, distributed within the \textsc{PseudoDojo} (\ourtable), using the new framework of the optimized norm-conserving Vanderbilt pseudopotential (\textsc{ONCVPSP}).~\cite{PhysRevB.88.085117,PhysRevB.95.239906} The main advantage of \textsc{ONCVPSP} is that it produces NCPPs that are usually softer, i.e. lead to converged results at lower cutoff energies, and more accurate (semi-core states can be included via multiple projectors) than traditional NCPPs. 
Moreover, \textsc{ONCVPSP} is interfaced with \textsc{libxc}~\cite{Marques20122272} and can therefore generate NCPPs for many XC flavors with or without spin-orbit (SO) terms. Our main goal is to provide a set of well-tested and accurate NCPPs that can be used for (a) applications in which the USPP/PAW formalism is not available or not implemented, (b) high-throughput calculations (HTC) and/or systematic studies involving NCPPs e.g. validation of a new PAW/USPP implementation or comparison of the accuracy of the different formalisms in different domains like NC+GW vs PAW+GW. See for example our recent systematic study on the convergence properties of $GW$.\cite{HTGW17}

The \textsc{PseudoDojo} is an open source project hosted on github and provides a user web-interface at pseudo-dojo.org. We provide pseudopotential files that can be used immediately, as well as the corresponding inputs so that users can tune or change some parameters (e.g. the XC functional) according to their needs. Moreover, we provide an open source python toolbox, that can be used for the automatized generation and validation of pseudopotentials. The pseudopotential files are available on the \textsc{PseudoDojo} web-interface in the \textsc{ABINIT} psp8 format, in the UPF2 format and in the PSML 1.1 XML format shared by \textsc{SIESTA} and \textsc{ABINIT}. The input files, the results of the generation, and the test results are presented via jupyter notebooks\cite{JN} as static HTML pages. Finally, each pseudopotential is linked to a \dreport\: file with a full record of the different tests that were performed to validate the pseudopotential (cutoff convergence, \df, GBRV tests~\cite{Garrity_2014}). One can hence easily compare PSPs for a given element and then select the most appropriate one according to a chosen criterion (e.g. efficiency vs accuracy).  

The remaining of this article is organized as follows: The \textsc{ONCVPSP} formalism and the most important differences with respect to standard NCPPs are discussed in section~\ref{sect:Formalism}. Subsequently the \textsc{PseudoDojo} project is presented in Section~\ref{sect:pseudodojo} including the python framework used for the automatic generation and validation of the pseudopotentials (PSPs) as well as the web interface that provides access to the PSPs. 
Section~\ref{sect:Pseudopotentials} describes the general strategy employed to generate the \ourtable. 
Sections~\ref{sect:convergence} and~\ref{sect:validation} describe the performance of the PSPs in convergence, \df,~\cite{Lejaeghere_2013} and GBRV~\cite{Garrity_2014} tests. A detailed discussion per group of elements of the choices made and the parameters employed for the pseudization is given in section~\ref{sect:discussion}.

\section{Formalism}\label{sect:Formalism}

The accuracy of the \textsc{ONCVPSP} pseudopotentials is based on the use 
of two projectors and generalized norm conservation to reproduce the binding 
and scattering properties of the all-electron potentials. The underlying 
formalism of generalized norm conservation was developed by Vanderbilt and 
used to generate ultrasoft pseudopotentials (USPPs)~\cite{Vanderbilt_1990}. Suppose we 
construct several radial pseudo-wavefunctions $\varphi_i$ at energies 
$\varepsilon_i$ and angular momentum $\ell$, which agree with all-electron 
radial wavefunctions $\psi_i$ outside a ``core radius" $r_c$, have 
continuous values and first derivatives at $r_c$, and satisfy
\begin{equation}\label{eq-1}
\left\langle\varphi_i\left|\vphantom{\psi_i\psi_j}\right.\varphi_j\right\rangle_{r_c} = 
\left\langle\psi_i\left|\vphantom{\psi_i\psi_j}\right.\psi_j \right\rangle_{r_c}
\end{equation}
where the notation indicates that norms and overlaps are calculated 
inside $r_c$.  These $\varphi_i$ obey generalized norm conservation 
in the sense that the integrated charge density inside $r_c$ of any linear 
combination of the $\varphi_i$ equals that of the corresponding 
combination of $\psi_i$.  Let these actually be $r$ times the radial 
wavefunctions so that the kinetic energy operator simplifies to 
$T = [ - d^2 / dr^2 + \ell (\ell  + 1)/ r^2 ] / 2 $ in atomic units.  We introduce the projectors
\begin{equation}\label{eq-2}
\left| \chi_i \right\rangle  = \left( \varepsilon_i - T - 
V_{\rm{loc}}\right)\left| \varphi_i \right\rangle ,
\end{equation}
where ${V_{{\rm{loc}}}}$ is a local potential agreeing with the all-electron 
potential outside ${r_c}$, and form the non-local operator
\begin{equation}\label{eq-3}
{V_{{\rm{NL}}}} = \sum\limits_{i,j} {\left| {{\chi _i}} \right\rangle 
{{({B^{ - 1}})}_{ij}}\left\langle {{\chi _j}} \right|} \,
\end{equation}
where
\begin{equation}\label{eq-4}
B_{ij} = \left\langle \varphi_i \left| \vphantom{\varphi_i\chi_j}\right. \chi_j \right\rangle .
\end{equation}
Generalized norm conservation is sufficient to prove that ${B_{ij}}$ is a 
symmetric matrix, so ${V_{{\rm{NL}}}}$ is a Hermitian operator.  Furthermore, 
for solutions of the non-local radial Schr{\"o}dinger equation
\begin{equation}\label{eq-5}
\left( {T + {V_{{\rm{loc}}}} + {V_{{\rm{NL}}}}} \right)\varphi  = 
\varepsilon \varphi  \,,
\end{equation}
$d\ln \varphi /dr$ and ${d^2}\ln \varphi /d\varepsilon dr$ will agree with 
those of all-electron solutions $\psi$ at each ${\varepsilon _i}$ for 
$r \ge {r_c}$ \cite{Vanderbilt_1990} In fact Eq.(\ref{eq-3}) is transformed  using the 
eigenvectors of ${B_{ij}}$ to form orthonormal projectors 
$\left| {{{\tilde \chi }_i}} \right\rangle $ for a computationally convenient 
diagonal ${V_{{\rm{NL}}}}$.  

It is straightforward to show that these 
principles apply to positive-energy scattering as well as bound-state 
solutions, paralleling the result for basic norm conservation.\cite{Hamann_1989}
The local potential ${V_{{\rm{loc}}}}$ is generally chosen to be a smooth 
polynomial continuation of the all-electron potential ${V_{{\rm{AE}}}}$ to 
the origin, continued from the smallest ${r_c}$ among the 
included $\ell$. This allows considerable flexibility which can sometimes 
be exploited to further extend the range of log-derivative agreement 
for one or more $\ell $.
Note that ultrasoft pseudopotentials are constructed from 
${\varphi _i}$ which do not satisfy Eq.~(\ref{eq-1}), but are compensated by the 
introduction of an overlap operator on the right side of the radial 
Schr{\"o}dinger equation and an augmentation contribution to the charge 
density~\cite{Vanderbilt_1990}.

The strategy employed in \textsc{ONCVPSP} to obtain the accuracy of 
two-projector ultrasoft potentials and nearly competitive convergence while 
retaining the simplicity of norm conservation is to enlist the convergence 
metric introduced by Rabe and coworkers (RRKJ).\cite{Rappe_1990} They observed that the error 
in the kinetic energy made by truncating the radial Fourier expansion of a 
pseudo-wavefunction $\varphi$ at some cutoff wave vector ${q_{\rm{c}}}$ was 
an accurate predictor of the convergence error made by similarly truncating 
the planewave expansion in calculations for solids. An optimization 
formalism was developed independently for \textsc{ONCVPSP} \cite{PhysRevB.88.085117,PhysRevB.95.239906}. The pseudo-wavefunction is first constrained to satisfy $M$ continuity 
constraints,
\begin{equation}\label{eq-6}
{\left. {\frac{{{d^n}\varphi }}{{d{r^n}}}} \right|_{{r_c}}} = 
{\left. {\frac{{{d^n}\psi }}{{d{r^n}}}} \right|_{{r_c}}}\,,\,n = 0,M - 1\,.
\end{equation}
$\varphi$ is then expanded in a set of $N \ge M + 3$ basis 
functions $\{ {\xi _i}\} ,$ initially chosen to be an orthogonalized set of 
spherical Bessel functions.  Employing singular-value analysis, a 
linear combination ${\varphi _0}$ is formed which satisfies Eq.~(\ref{eq-6}), as well as a new set 
of $N - M$ ``null space" basis functions $\{ \xi _i^{\rm{N}}\} ,$ which 
are mutually orthonormal, orthogonal to ${\varphi_0}$ , and give zero 
contribution to Eq~(\ref{eq-6}) when added to ${\varphi_0}$.  A generalized 
residual kinetic energy operator is defined as:
\begin{equation}\label{eq-7}
\left\langle {{\xi _i}} \right|{\hat E^{\rm{r}}}({q_{\rm{c}}})\left| 
{{\xi _j}} \right\rangle  \equiv \int_{{q_{\rm{c}}}}^\infty  
{{\xi _i}(q){\xi _j}(q){q^4}dq} \,\,
\end{equation}
using the radial Fourier transform 
\begin{equation}\label{eq-8}
{\xi _i}(q) = 4\pi \int_0^\infty  {{j_\ell }(qr){\xi _i}(r){r^2}dr} \,.
\end{equation}
The cutoff energy error to be minimized for optimum convergence ${\left\langle
 \varphi  \right|{\hat E^{\rm{r}}}\left| \varphi  \right\rangle}$ can now 
be expressed as
\begin{equation}\label{eq-9}
{E^{\rm{r}}}({q_c}) = \left\langle {{\varphi _0}} \right|{\hat E^{\rm{r}}}
\left| {{\varphi _0}} \right\rangle  + 2\sum\limits_{i = 1}^{N - M} 
{{y_i}\left\langle {{\varphi _0}} \right|{{\hat E}^{\rm{r}}}\left| 
{\xi _i^{\rm{N}}} \right\rangle }  + \sum\limits_{i,j = 1}^{N - M} 
{{y_i}{y_j}\left\langle {\xi _i^{\rm{N}}} \right|{{\hat E}^{\rm{r}}}\left| 
{\xi _j^{\rm{N}}} \right\rangle }
\end{equation}
where ${y_i}$ are the coefficients of the $\xi _i^{\rm{N}}$ basis functions 
to be added to ${\varphi _0}$.  The ${y_i}$ are subject to the norm constraint
\begin{equation}\label{eq-10}
\sum\limits_{i = 1}^{N - M} {y_i^2}  = \left\langle {\psi }
 \left | \vphantom{\varphi_0} \right. 
 \psi  \right\rangle_{r_c} - \left\langle \varphi_0
 \left | \vphantom{\varphi_0}
 \right.
 \varphi_0 \right\rangle_{r_c}\,.
\end{equation}
Standard methods for minimizing Eq.~(\ref{eq-9}) subject to Eq.~(\ref{eq-10}) can be quite unstable.  
Instead, the positive-definite $E_{ij}^{\rm{r}}$ matrix, 
the last term in Eq.~(\ref{eq-9}), is diagonalized finding its eigenvalues ${e_i}$ and using
its eigenvectors to form the new "residual" basis function set
$\{ \xi _i^R \}$ as linear combinations of the $\xi _i^N$.
When these functions are added to ${\phi_0}$ with coefficients ${x_i}$ to form ${\varphi}$, 
the residual energy takes the diagonal quadratic form
\begin{equation}\label{eq-11}
{E^{\rm{r}}} = E_{00}^{\rm{r}} + \sum\limits_{i = 1}^{N - M} {\left( 2{f_i}{x_i}\, + \,{e_i}{x_i^2} \right)}\,.
\end{equation}
where ${f_i} = \left\langle {{\varphi _0}} \right|{{\hat E}^{\rm{r}}}\left| 
{\xi _i^{\rm{R}}} \right\rangle$.
The ${x_{_i}}$  satisfy the same norm constraint as the  ${y_{_i}}$ in Eq.~(\ref{eq-10}).
The ${e_i}$ span a very large dynamic range $\sim{10^6} - {10^8}$, which may explain the difficulties 
in applying standard optimization procedures to Eq.~(\ref{eq-9}).  We next solve the constraint equation
for ${x_1}$, the coefficient corresponding to the smallest ${e_i}$, 
as a functiion of ${x_{2,}},...\,,{x_{N - M}}$:
\begin{equation}\label{eq-12}
 {x_1} = s\left[ \left\langle {\psi }
 \left | \vphantom{\varphi_0} \right. 
 \psi  \right\rangle_{r_c} - \left\langle \varphi_0
 \left | \vphantom{\varphi_0}
 \right.
 \varphi_0 \right\rangle_{r_c} - \sum\limits_{i = 2}^{N - M} {x_i^2}  \right]^{1/2}  \,.
\end{equation}
 Its sign $s$ determined by the requirement 
that ${f_1}{x_1}$ be negative at the minimum.  Setting the derivatives of ${E^{\rm{r}}}$ 
with respect to  ${x_{2,}},...\,,{x_{N - M}}$ to zero using Eq.~(\ref{eq-12}) for ${x_1}$ we find
\begin{equation}\label{eq-13}
 {x_i} = {{ - {f_i}} / {\left( {e_i} - {e_1} - {{f_1}/{x_1}} \right)}} \,.
\end{equation}
The denominator in Eq.~(\ref{eq-13}) is always positive, so the sum in Eq.~(\ref{eq-12}) is a monotonically
increasing function of $|{x_1}|$ starting from zero for $|{x_1}|=0$, and Eq.~(\ref{eq-12}) can be solved by a straightforward interval-halving search on ${|x_1|}$.\cite{PhysRevB.95.239906} 
The optimum ${x_i}$ are based on a prescribed ${q_c}$. However, Eq.~(\ref{eq-9}) can be evaluated for any cutoff $q$ using ${y_i}$ calculated from the ${q_c}$-optimized ${x_i}$, thereby providing a kinetic-energy-error per electron convergence profile.  

The above procedure is applied to the first (lowest energy) projector ${\varphi _1}$ in the two-projector 
generalized norm-conserving construction.  For the second projector, the convergence-optimized ${\varphi _1}$ is used to add the linear 
$\left\langle\varphi _1\left|\vphantom{\varphi _1\varphi _2}\right.\varphi _2\right\rangle_{r_c}$ overlap constraint to the 
continuity constraints of Eq.~(\ref{eq-6}). The procedure continues as above, retaining the original spherical-Bessel-function basis set for convenience, and the 
coefficients are found determining the convergence-optimized ${\varphi _2}$.  While 
there are only $N - M - 1$ degrees of freedom for norm conservation and 
optimization, convergence profiles are usually quite comparable to those 
for ${\varphi _1}$.  As the broad range of ${\hat E^{\rm{r}}}$ eigenvalues 
suggests, convergence improvements decrease rapidly as more degrees of freedom 
are added, and 3-5 invariably suffice.

While it is observed that scattering states can be used as well as 
bound states to satisfy the generalized norm-conservation requirements and 
retain its resulting accuracy, they cannot be used in the optimization because 
the radial Fourier transform of such a $\varphi $ is essentially a delta 
function of $q$.  To deal with this, an artificial all-electron 
bound state is created at each positive ${\varepsilon _i}$ by adding a smoothly rising 
barrier to the all-electron potential beginning at ${r_c}$.  A satisfactory 
form is
\begin{equation}\label{eq-14}
{V_{{\rm{AEB}}}}(r) = {V_{{\rm{AE}}}}(r)\, + \,{v_\infty }
\theta (x){x^3}/(1 + {x^3})\;;\;x = (r - {r_c})/{r_{\rm{b}}}\,,
\end{equation}
where the height and shape parameters ${v_\infty }$ and ${r_{\rm{b}}}$ are 
chosen to bind a state with the appropriate number of nodes 
at ${\varepsilon _i}$ and produce a decaying tail roughly comparable to 
those of the highest occupied bound states.  The optimized convergence 
properties of the corresponding bound pseudo-wavefunctions are typically 
comparable to those of the valence functions \cite{PhysRevB.88.085117,PhysRevB.95.239906}.

The symmetry of ${B_{ij}}$ and other consequences of general norm conservation 
are strictly true for pseudopotentials based on non-relativistic all-electron 
calculations.  Nonetheless, we have proceeded to apply them to 
scalar-relativistic~\cite{koelling97} and fully relativistic calculations.  
In practice, an asymmetry of $\sim{10^{ - 4}}$ to 
${10^{ - 5}}$ was found for both light and heavy atoms, so ${B_{ij}}$ was simply symmetrized before proceeding. This manifests itself in disagreements of comparable 
magnitude in comparisons of quantities such as eigenvalues and norms computed 
with the final pseudopotentials. In the fully relativistic case, the large 
component of the Dirac wavefunction is renormalized and only it is used to 
compute the Eq.~(\ref{eq-1}) norms and overlaps and the matching constraints of Eq.~(\ref{eq-6}).  
This yields errors comparable to the scalar-relativistic case, and an order 
of magnitude smaller than obtained using both components.

Relativistic non-local pseudopotentials are generated as sums 
over total angular momenta
$j = \ell  \pm 1/2 ,j > 0$, of terms
$V_j^{\rm{Rel}}(\bf{r},\bf{r'})$ like Eq.~(\ref{eq-3}).  While these may be used 
directly in some applications, most require potentials in the (schematic) form
\begin{equation}\label{eq-15}
V({\bf{r}},{\bf{r'}}) = {V_{{\rm{loc}}}} + \sum\limits_\ell  
{\left[ {V_\ell ^{{\rm{SR}}}({\bf{r}},{\bf{r'}})\; + \;{\bf{L}} \cdot 
{\bf{S}}V_\ell ^{{\rm{SO}}}({\bf{r}},{\bf{r'}})} \right]} \,,
\end{equation}
where
\begin{equation}\label{eq-16}
V_{\ell}^{\rm{SR}} = \frac{(\ell  + 1)V_{\ell  + 1/2}^{\rm{Rel}} + 
\ell V_{\ell - 1/2}^{\rm{Rel}}}{2\ell +1},
\;\;V_\ell^{\rm{SO}} = \frac{2\left(V_{\ell  + 1/2}^{\rm{Rel}} - 
V_{\ell - 1/2}^{\rm{Rel}}\right)}{2\ell +1}
\end{equation}

Direct use of these ``scalar-relativistic" and ``spin-orbit" potentials 
as sums and differences is both cumbersome and requires subtractions of 
many nearly equal quantities in applications, with the resulting 
inaccuracies.  For these applications, \textsc{ONCVPSP} forms new projectors 
$\left| {\tilde \chi _\ell ^{{\rm{SR}}}} \right\rangle $ and 
$\left| {\tilde \chi _\ell ^{{\rm{SO}}}} \right\rangle $ from their 
eigenfunctions to create diagonal non-local operators, some of whose 
eigenvalue coefficients are negligibly small.  Either form can be selected.

\section{The \textsc{PseudoDojo}}\label{sect:pseudodojo}

\subsection{The \textsc{PseudoDojo} python framework}

The \textsc{PseudoDojo} is a python framework for the automatic generation and validation of pseudopotentials. It consists of three different parts: (1) a database of reference results produced with AE and PSP codes, (2) a set of tools and graphical interfaces that facilitate the generation and the initial validation of the PSPs and (3) a set of scripts to automate the execution of the different tests in a crystalline environment (automatic generation of input files, job submission on massively parallel architectures, post-processing and analysis of the final results).

The database currently contains the reference all-electron results for the \df\: and the GBRV benchmarks as well as the structural parameters used in these tests. The \textsc{PseudoDojo} is presently interfaced with \textsc{ONCVPSP}. It provides a GUI to set up the input parameters and visualize the results of the comparison of the PSP to the atomic reference calculation, e.g. their logarithmic derivatives. In particular, series of PSPs can be generated for ranges of input parameters. Finally, after the initial 'internal' validation against the atomic reference calculation the implemented 'external' tests can be executed via \textsc{AbiPy} and \textsc{ABINIT}.~\cite{Gonze2009,Gonze2016} The currently implemented external tests include the \df, the GBRV tests, automatic convergence testing the evaluation of the acoustic modes at $\Gamma$ within DFPT, and ghost state testing of the electronic structure up to high energies ($\sim$ 200~eV above the Fermi level). All of these can be executed fully automatically on various parallel architectures. Interface to other DFT codes, additional tests, and other pseudopotential generators can be easily added.

\subsection{The Dojo-report}

An important aspect of the \textsc{PseudoDojo} is keeping track of the results of various validation tests. To this end, the \textsc{PeudoDojo} creates a report for each pseudopotential. This \dreport~is a human-readable text document in JSON format,\footnote{JSON (Java-Script object notation) is a language-independent data format that uses text to represent objects in the form of lists and attribute-value pairs. We decided to use JSON to store our data because code for parsing and generating JSON data is readily available in many programming languages (python provides native support for JSON in the python standard library). Besides it can be used to transmit data between a server and web application, as an alternative to XML.}\selectlanguage{english} containing entries for each test. It is automatically produced by the python code at the end of the test. In addition to the raw data it contains the final results as function of $E_\mathrm{c}$.

The data in the report is in principle not intended for the {\em ab-initio} code.\footnote{An exception may eventually be the direct use of the hints on the cutoff energy. Currently, however, there is no specification for this field neither in the \textsc{ABINIT} format nor in the UPF one.} The main goal of the \dreport\: is to keep a record of the different tests, so that it can be used by high-level languages (e.g. python) to read the data and produce plots or rank pseudopotentials associated to the same element according to some criterion. In addition, the information in the \dreport\: can be used to set up high-throughput calculations. Finally, new validation tests can be easily added to the JSON document. 

\subsection{The \textsc{PseudoDojo} web interface}

In addition to the \textsc{PseudoDojo} python framework itself, the \textsc{PseudoDojo} provides a web-interface~\cite{pdwi} for the on-line visualization of both the internal and external validations. The web-interface allows for a fast visualization of the test results for a particular pseudopotential, via the HTML version of the DojoReport generate automatically from a Jupyter Notebook,\cite{JN} without having to install the python package. Both the pseudopotential files and the corresponding input files can be downloaded. 

\section{The PD-PBE tables}\label{sect:Pseudopotentials}

\subsection{General design principles}

Despite several significant improvements proposed in the literature,~\cite{Troullier_1991, Rappe_1990, Hamann_1989, PhysRevB.88.085117,PhysRevB.95.239906} elements with localized $d$- or $f$-electrons are still difficult to pseudize within the NC formalism. For this reason, unlike other similar projects, {\em e.g.} the GBRV table in which all the ultrasoft pseudopotentials require an $E_\mathrm{c}$ less than 20 Ha~\cite{Garrity_2014} or the SG15 table,~\cite{Schlipf2015} which is mainly focusing on efficiency, we do not make any attempt to generate an entire periodic table of NCPPs that converge below the {\em same} $E_\mathrm{c}$. Instead, we mainly focus on accuracy and transferability and attempt to tune the pseudization parameters so that elements with similar electronic configurations require similar $E_\mathrm{c}$ to achieve convergence. 

In this first version of the \textsc{PseudoDojo} we present and discuss the pseudopotentials for the GGA-PBE exchange correlation (XC) functional.~\cite{Perdew1996} For this functional well-tested sets of reference data are available. Pseudopotentials for the LDA-PW~\cite{Perdew92} and PBEsol~\cite{Perdew08} functionals are also available via the \textsc{PseudoDojo} web interface, reference values for these functionals are currently under development. Other XC functionals can be generated easily, especially since as of version 3.0 the \textsc{ONCVPSP} package is interfaced with the libxc library enabling well over 250 XC functionals.~\cite{Marques_2012} For each flavor of exchange-correlation functional we define a standard and a stringent accuracy version.

For those elements in which the separation between core and valence is not obvious, we provide a version with and without semi-core electrons. As a rule of thumb, NCPPs with semi-core states are more accurate and transferable since the error introduced by the frozen-core approximation is reduced. Moreover, semi-core states may be needed for accurate \textit{GW} calculations, in particular in those systems in which there is an important overlap between valence and semi-core electrons and therefore a significant contribution to the exchange part of the self-energy.~\cite{Scherpelz2016, PhysRevB.69.125212} We adapt the notation, e.g. Fe-sp, to indicate additional semi-core states included in the valence. 

For elements that show a particularly slow convergence in reciprocal space (e.g. transition metals) we also provide two different versions: normal and high. The default version, normal accuracy, is designed to give a good description of the scattering properties of the atom in different chemical environments with a reasonable $E_\mathrm{c}$. The high-accuracy version, with small core radii, requires a larger $E_\mathrm{c}$ to converge but is more transferable and can be used for accurate first-principles calculations or for the study of systems under high pressure. The high accuracy version is also recommended for calculations in magnetic systems.

In special cases, discussed in section~\ref{sect:discussion}, we also provide low accuracy pseudopotentials. We do this when the standard version converges only at cutoff energies higher than 40~Ha. 

Except for some noticeable exceptions listed in Table~\ref{tab:nproj}, all the PSPs of our tables contain two projectors per angular channel. This ensures a logarithmic derivative in close agreement with the AE counterpart up to at least 3-5 Ha. In many cases, we achieve agreement even up to 10 Ha. Further element specific details will be discussed in section~\ref{sect:discussion}. 

\begin{table}
\center
\begin{tabular}{lcccc}
\toprule
Pseudo  & $s$ & $p$ & $d$ & $f$\\
\midrule
H           & 2 & 1 & 0 & 0\\
He          & 2 & 1 & 0 & 0\\
O-high      & 2 & 2 & 1 & 0\\
O           & 2 & 2 & 1 & 0\\
F           & 2 & 2 & 1 & 0\\
Lanthanides & 2 & 2 & 2 & 1\\
Au-sp       & 2 & 2 & 2 & 1\\
Hg-sp       & 2 & 2 & 2 & 1\\
\bottomrule
\end{tabular}
\caption{Number of projectors in the $s$, $p$, $d$, and $f$ channels. All other pseudopotentials are constructed using two projectors per angular channel. The highest $l$ projector is $p$ of H-Mg (except for F and O where it is $d$), $d$ for Al-Xe and Tl-Rn, and $f$ for Cs-Hg except for Ba where it is $d$.}\selectlanguage{english} \label{tab:nproj}
\end{table}

In general, we enforce the continuity of the derivatives of the pseudized potentials at $r_c$ up to the fourth order ($M$ in eqn~\ref{eq-6}, input parameter \texttt{ncon=4}). This is done in order to avoid possible problems in the computation of elastic properties introduced by the RRKJ optimization technique (see also the discussion in Ref.~\cite{PhysRevB.88.085117,PhysRevB.95.239906}). Those pseudopotentials that deviate from this rule are listed in Table~\ref{tab:ncon} and discussed in more detail in section~\ref{sect:discussion}. 
A drawback of this additional requirement is that it usually leads to pseudopotentials that are slightly harder than the ones obtained by enforcing continuity up to the third order as it is commonly done. In general, we found that one can decrease the required $E_\mathrm{c}$ by $\sim5$~Ha if \texttt{ncon=3} (continuous derivatives up to third order) is used. 

\begin{table}
\center
\begin{tabular}{lcccc}
\toprule
Pseudo     & $s$ & $p$ & $d$\\
\midrule
In-spd     & 5 & 5 & 4\\
In-d       & 5 & 5 & 4\\
Ga-low     & 3 & 3 & 3\\
Fe-sp      & 3 & 3 & 3\\
Fe-sp-high & 3 & 3 & 3\\
\bottomrule
\end{tabular}
\caption{Order of the derivative of the pseudized potential that is still continuous. Only those pseudopotentials are listed that deviate from having continuity up to exactly the fourth-order derivative at the core radius for each angular channel.}\selectlanguage{english} \label{tab:ncon}
\end{table}

It is well known that nonlinear core corrections (NLCC) improve the transferability of pseudopotentials.~\cite{PhysRevB.26.1738} PSPs that do not include semi-core states usually improve the most. However, even when semi-core states are present, adding NLCCs has benefits. They remove the nonphysical oscillations of the local part close to the origin, oscillations which often appear in the case of gradient-corrected functionals when the total local potential is unscreened.
These oscillations create problems if the potentials are represented in a non-planewave basis sets but also tend to spoil convergence in Fourier and real space.

In \ourtable, a NLCC is included in all the PSPs with electrons frozen in the core except for the third row semi-core PSPs (Na-sp -- Cl-sp) and Ne. We use a recently implemented NLCC following Teter, which contains two parameters.~\cite{PhysRevB.48.5031} These model core charges are by construction smooth in both real and reciprocal space, which significantly improves convergence. Teter suggested to use these two parameters to minimize the difference between the chemical hardness of the pseudo and the AE wavefunction.~\cite{PhysRevB.48.5031} In constructing \ourtable, however, we did not observe a clear correlation between the PSP quality (in reproducing AE results for crystalline test systems) and the level at which the pseudized wavefunction reproduces the AE chemical hardness. 
Teter's approach, on the other hand, revealed to be quite successful in the case of elements with localized AE core charges.
Standard models, indeed, produce charges that are either too peaked and thus difficult to integrate on a homogeneous mesh in real-space or model charges with strong oscillations in the high-order derivatives required for DFPT calculations.
This can spoil the convergence of the physical properties with the
cutoff energy and have disastrous effects for density functional perturbation theory calculations, in particular for the fulfillment of the acoustic sum rule. 
This is the reason why we add a test in the \textsc{PseudoDojo} for the acoustic modes at $\Gamma$. Large deviations from zero (when the ASR is not enforced by the code) usually indicate that the model core charge and its derivatives cannot be correctly described with a sufficiently small cutoff energy and these inaccuracies will likely affect the phonon modes at other q-points as well.

\section{Convergence and energy cutoff hints}
\label{sect:convergence}

The different options described in the previous section lead for most elements to several PSPs. To assist users in selecting pseudopotentials, we define two tables: standard and stringent accuracy, both of which contain only one PSP per element. For about half the elements, the stringent table contains a different, more accurate, PSP than the standard table. In this and the next section, we evaluate the results of the convergence studies and the validation tests for these two tables. 

The design of the \ourtable\: allows for different required energy cutoffs ($E_\mathrm{c}$) for each pseudopotential. Moreover, different physical properties usually show a different convergence behavior with respect to $E_\mathrm{c}$. 
Typical examples are phonons and the bulk modulus, which are much more sensitive to the truncation of the PW basis-set than, e.g., the total energy. It is however useful to have an initial estimate for the starting $E_\mathrm{c}$ for the convergence study, both for 'normal' users and for high-throughput calculations (HTC). We therefore provide calculated high, normal, and low precision hints for $E_\mathrm{c}$: $E_\mathrm{c}^\mathrm{h,n,l}$  based on different tests.\footnote{We decided to introduce three different levels of precision because one can use this information to implement automatic HTC workflows that are both efficient and reliable. For example, one can perform an initial HTC screening on many systems with the low precision $E_\mathrm{c}$ in order to select the most promising candidates and then refine the search with calculations done with the normal or the high precision $E_\mathrm{c}$. In the same spirit, one can implement machine-friendly convergence studies in which the convergence of the physical property of interest is validated without any human intervention by just analyzing the difference between a calculation done with the ``normal'' setup and the one performed with the high-precision version.}\selectlanguage{english}

The \textsc{ONCVPSP} code already provides initial hints for $E\mathrm{c}$ based on the convergence of the electronic eigenvalues in the atomic environments ($\epsilon$). We used these values to define an initial mesh of $E_\mathrm{c}$ values (a dense sub-mesh with a step of 2 Ha around the initial value provided by the PSP generator continued by a coarse mesh with a step of 10 Ha to ensure absolute convergence). On this mesh, we use the \textsc{PseudoDojo} framework to compute the \df, the GBRV parameters, and the phonons at $\Gamma$ as a function of $E_\mathrm{c}$. The final results as well as the total energies used for fitting the EOS curve are all saved in the \dreport. 

The hints are calculated according to the parameters specified in Table~\ref{tab:hint-pars}. Using the hint for one of the accuracies ensures that the absolute value of the indicated quantity is smaller than the indicated bound. $O^c$ indicates the converged value of observable $O$, which is obtained from the largest $E_\mathrm{c}$ grid point. This point is initially 22 Ha higher than the high precision estimate given by \textsc{ONCVPSP}. All curves are however inspected manually to ensure convergence. In an automatic fashion, additional grid-points are added until a curve is approved with a converged tail.\footnote{Again, the additional points are stored in the \dreport. A fully automatic evaluation of the degree of convergence turned out to be too optimistic. Especially quantities like the \df\: turned out to occasionally have an oscillating behavior necessitating human inspection.}\selectlanguage{english} The hints are reported in the \dreport\: of each PSP file and listed in the supplementary material. Fig.~\ref{violin_cutoff_hints} summarizes the hints for the high and standard tables,\footnote{We use violin plots generated with the \textsc{Matplotlib} and \textsc{Seaborn} python packages to compare the different distributions of the test results. We use the Scott method to compute the kernel bandwidth and cut the plots off at the extremal values. Full details on the generation are included in the supplementary material} Tables~\ref{tbl_standard_hints} and~\ref{tbl_stringent_hints} report the statistics on the two tables.

\begin{table}
\center
\begin{tabular}{lcrcl}
\toprule
Observable & unit & low & normal & high \\
\midrule
$\epsilon - \epsilon_{AE}$ &(mHa/electron)   & --    &  $<1$ &  $<1$ \\
$\Delta_1 - \Delta_1^c $   &(meV)     &  $<2$ &  $<1$ &  $<0.5$ \\
TE - TE$^c$                &(meV/atom)& $<10$ &  $<5$ &  $<2$   \\
\bottomrule
\end{tabular}
\caption{Criteria for the low, normal and high hints of \ourtable. $\epsilon$ indicates the maximal deviation among electronic energy (not used for the low hint criterion), $\Delta_1$ the revised \df~as introduced in Ref~\cite{Jollet_2014}. TE indicates the total electronic energy per atom obtained at the equilibrium volume defined in the reference equilibrium structure as given the \df ~benchmark.}\selectlanguage{english}
\label{tab:hint-pars}
\end{table}

\makefigure{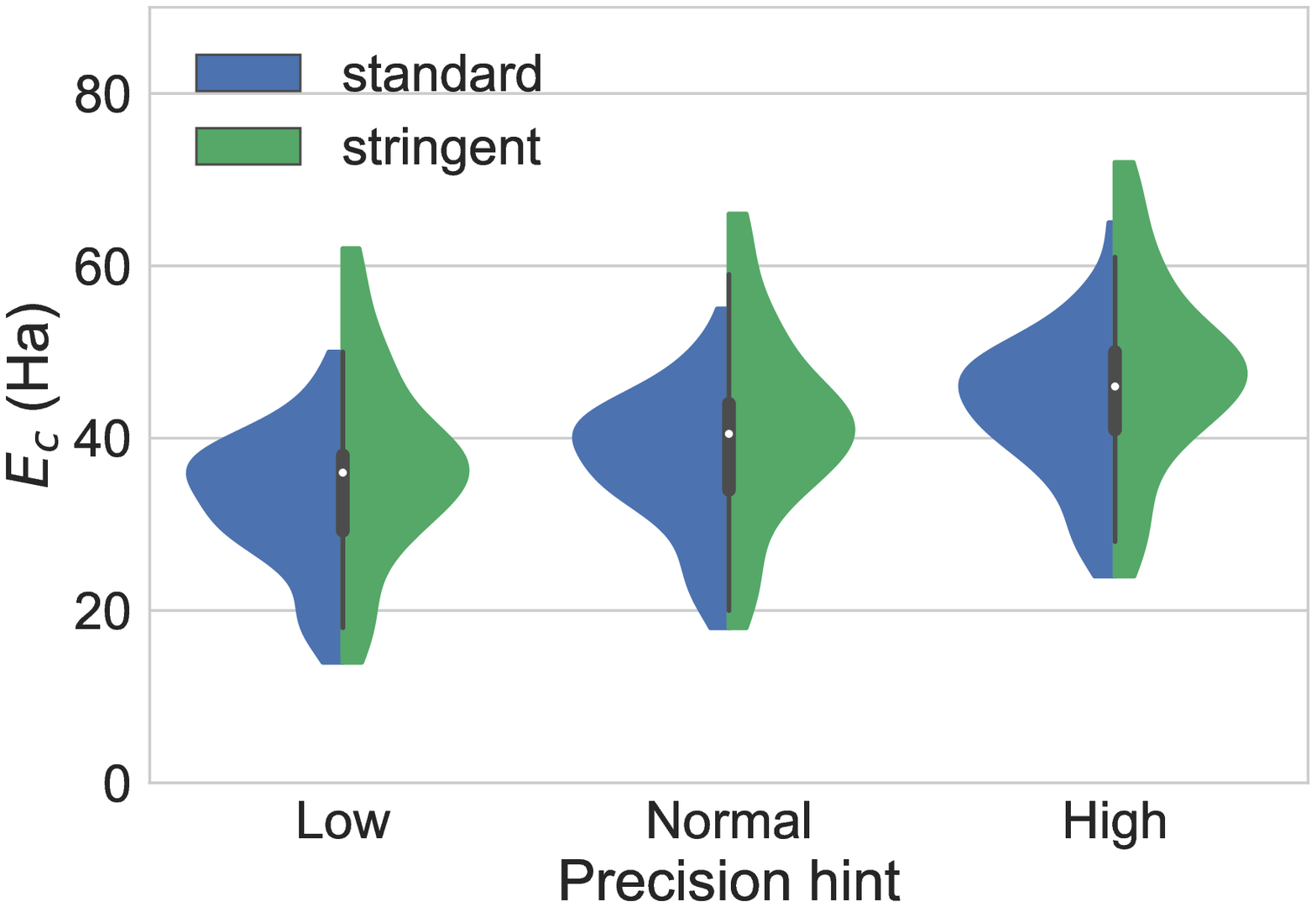}{Violin plot of the hints for the standard and high tables.}

\begin{table}\center\begin{tabular}{lrrrrr}
\toprule
{} &  $Z_\mathrm{val}$ &  $l_\mathrm{max}$ &  $E_\mathrm{c}^\mathrm{l}$ &  $E_\mathrm{c}^\mathrm{n}$ &  $E_\mathrm{c}^\mathrm{h}$ \\
\midrule
count &             72.00 &             72.00 &                      72.00 &                      72.00 &                      72.00 \\
mean  &             12.00 &              2.03 &                      32.74 &                      37.25 &                      43.36 \\
std   &              5.24 &              0.56 &                       7.69 &                       7.77 &                       8.13 \\
min   &              1.00 &              1.00 &                      14.00 &                      18.00 &                      24.00 \\
25\%  &              8.00 &              2.00 &                      28.75 &                      33.00 &                      38.75 \\
50\%  &             13.00 &              2.00 &                      33.50 &                      38.00 &                      44.00 \\
75\%  &             16.00 &              2.00 &                      38.00 &                      42.00 &                      48.25 \\
max   &             25.00 &              3.00 &                      50.00 &                      55.00 &                      65.00 \\
\bottomrule
\end{tabular}
\caption{Statistics on the low, normal and high hints for the standard table.}\label{tbl_standard_hints}\end{table}\selectlanguage{english}

\begin{table}\center\begin{tabular}{lrrrrr}
\toprule
{} &  $Z_\mathrm{val}$ &  $l_\mathrm{max}$ &  $E_\mathrm{c}^\mathrm{l}$ &  $E_\mathrm{c}^\mathrm{n}$ &  $E_\mathrm{c}^\mathrm{h}$ \\
\midrule
count &             70.00 &             70.00 &                      70.00 &                      70.00 &                      70.00 \\
mean  &             13.79 &              2.03 &                      37.19 &                      41.77 &                      47.83 \\
std   &              6.82 &              0.56 &                      10.80 &                      10.72 &                      10.80 \\
min   &              1.00 &              1.00 &                      14.00 &                      18.00 &                      24.00 \\
25\%  &              8.25 &              2.00 &                      31.25 &                      36.00 &                      42.00 \\
50\%  &             14.00 &              2.00 &                      37.00 &                      42.00 &                      48.00 \\
75\%  &             19.00 &              2.00 &                      43.50 &                      47.50 &                      52.00 \\
max   &             27.00 &              3.00 &                      62.00 &                      66.00 &                      72.00 \\
\bottomrule
\end{tabular}
\caption{Statistics on the low, normal and high hints for the stringent table.}\label{tbl_stringent_hints}\end{table}\selectlanguage{english}

\section{Discussion of the validation per table}
\label{sect:validation}

\subsection[]{$\Delta$-Gauge}

The \df\: is defined as the integral over the difference between the equation of state curve calculated using two different computational approaches within a predefined volume range expressed in meV per atom.\cite{Lejaeghere_2013} The physical quantities that are related to the \df\: are the parameters of the Birch-Murnaghan equation of state: the equilibrium volume $V_0$, the bulk modulus B, and the first derivative of the bulk modulus $B_1$. It was introduced by Cottenier and coworkers in 2014 and presently already 24 data sets have been calculated. This large number of data sets, involving 13 different codes (including 5 AE codes), makes the \df\: very useful in the validation of PSPs.~\cite{Lejaeghere2016} To test a PSP, one compares the results calculated using a PSP with those calculated using a reference AE code. The \df\: averaged over the periodic table between the most reliable AE data sets is around 0.3 - 0.5 meV. In this work we use Wien2k results as a reference.~\cite{Lejaeghere2016} The average delta of the NCPP tables with respect to the Wien2k results is 1.4.\footnote{1.4 is the average excluding the old FHI table, which has an average \df\: of 13.4. This value is however strongly dominated by a few elements, and hence not representative.} The drawbacks of the \df\: are however that the prescribed computational settings for the calculation are rather stringent making it unsuitable for fast pre-testing. Moreover, only single element compounds are included and only ground state properties are tested.

\makefigure{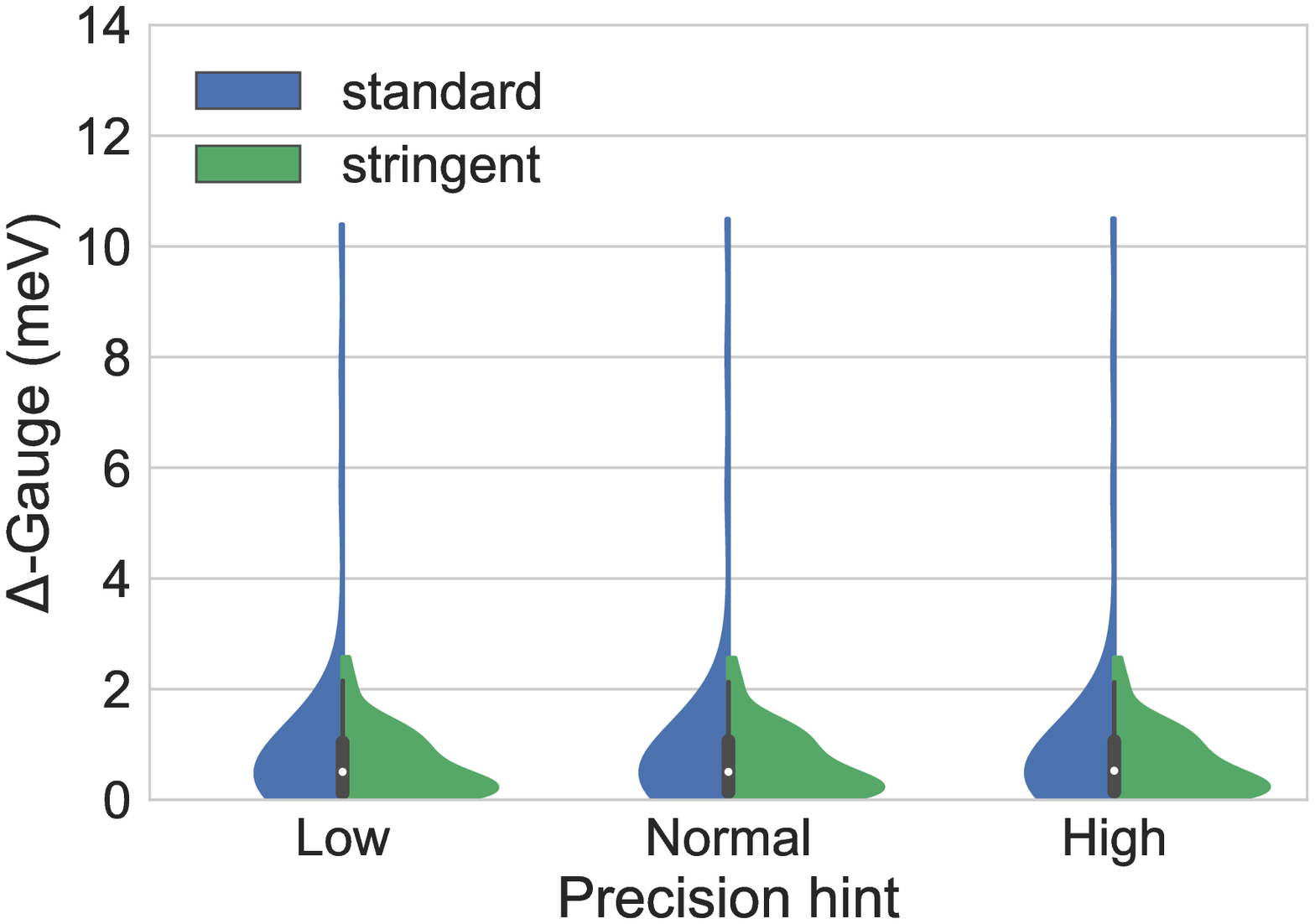}{Violin plot of the distribution of $\Delta$ values calculated at the low, normal, and high hints. The outliers, occurring at about 10.4, 7.7, and 5.9 for each $E_\mathrm{c}$ hint in the standard table, are Cr, Mn, and Fe respectively.}\selectlanguage{english}

Figure~\ref{violin_delta} summarizes the results of the \df\: tests for the standard and stringent tables. A full table of the results per pseudo is available in the supplementary information. For the \df\: test, the most significant difference between the two tables is confined to three elements: Cr, Mn, and Fe. For these elements the structures used in the \df\: test are magnetic. To resolve the magnetic structure a harder PSP is needed. When we exclude these three elements the mean \df\: are 0.70 and 0.64 for the standard and high tables.

\makefigure{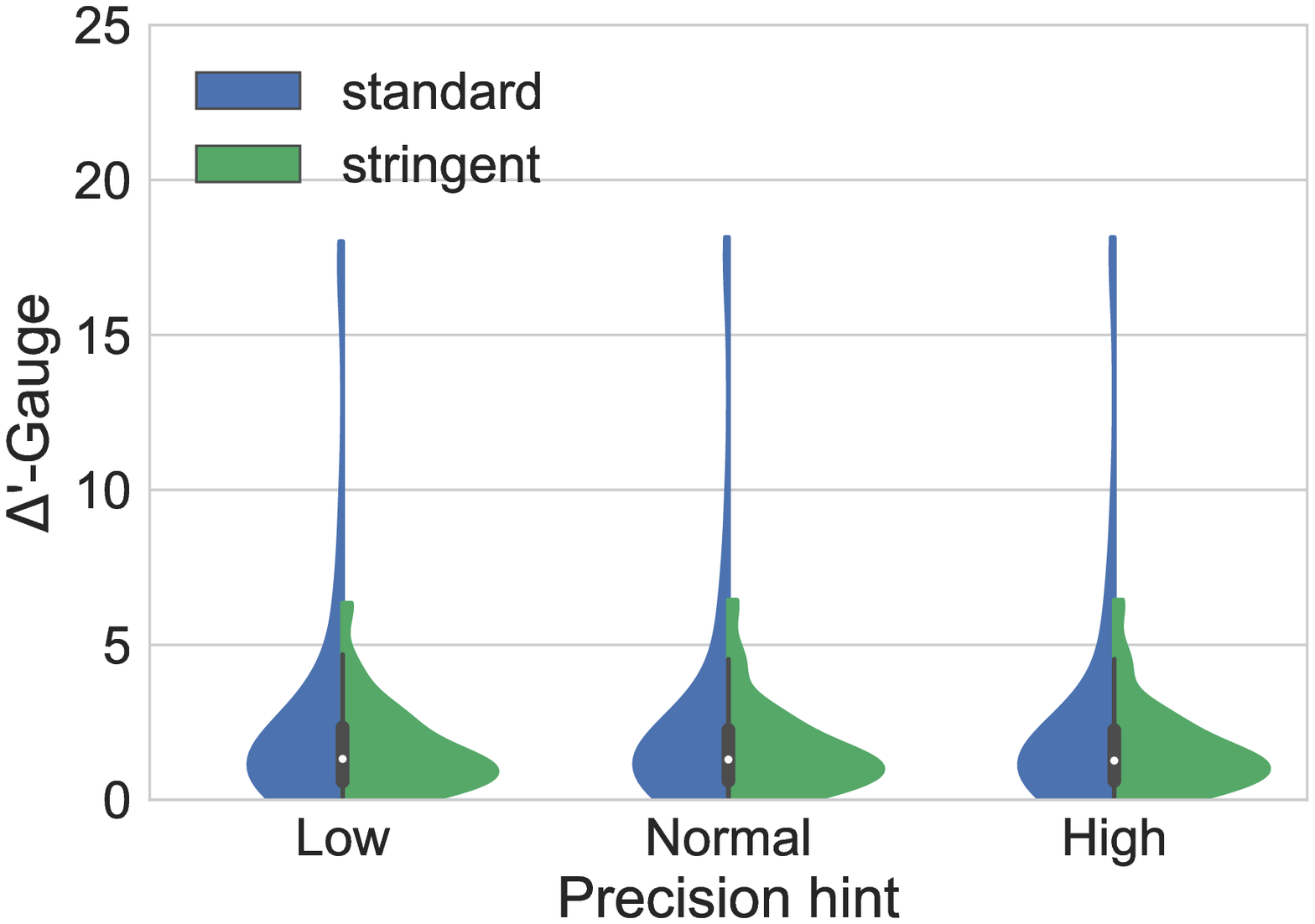}{Violin plot of the distributions \dfp\: values calculated at the low, normal, and high hints. Again Cr, Mn, and Fe are outliers in the standard table. In addition Hg (6.4), Sr (6.1), and Ba (5.0) appear as outliers also in the stringent table.}\selectlanguage{english}

The design of the \df\: is such that elements for which the bulk modulus is very soft are hard to test. The noble gas solids for instance always have a low \df. To remedy this, Jollet \textrm{et al.} have introduced a renormalized version of the \df\:: the \dfp.\cite{Jollet_2014} For the latter, a value less than 2 in general indicates an accurate potential for ground state structural properties.  Figure~\ref{violin_delta_prime} summarizes the results of the \dfp\: tests for the standard and high tables. In addition to what we have learned from the \df\:, the \dfp\: shows that Hg, Sr, and Ba are problematic elements. Their \dfp\: values, 7.2 (6.4), 6.1 (6.2), and 5.0 (4.8) respectively (stringent in brackets), are relatively high. We did not manage to create high accuracy versions that have a significantly better \dfp\: without becoming prohibitively expensive.

For both the high and low tables, Figs~\ref{violin_delta} and \ref{violin_delta_prime} indicate that the low and high hints already result in a converged \df\: and \dfp. This is made clearer in Fig.~\ref{violin_low_normals_errors}, which show the errors at low and normal $E_\mathrm{c}$ hint with respect to their converged values at the high hint. A similar convergence is observed for the equilibrium volume $V_0$. The outlier in $V_0$ is Ne. Since the bulk modulus of the solid state structure of Ne is however very small, the equation of state curve is very flat. This is a generic feature for all the crystal cases where the energy landscape is flat. This is thus not a problem of PSP but of the system. As a result, an error in $V_0$ does not affect the \df\: significantly. 

\makefigure{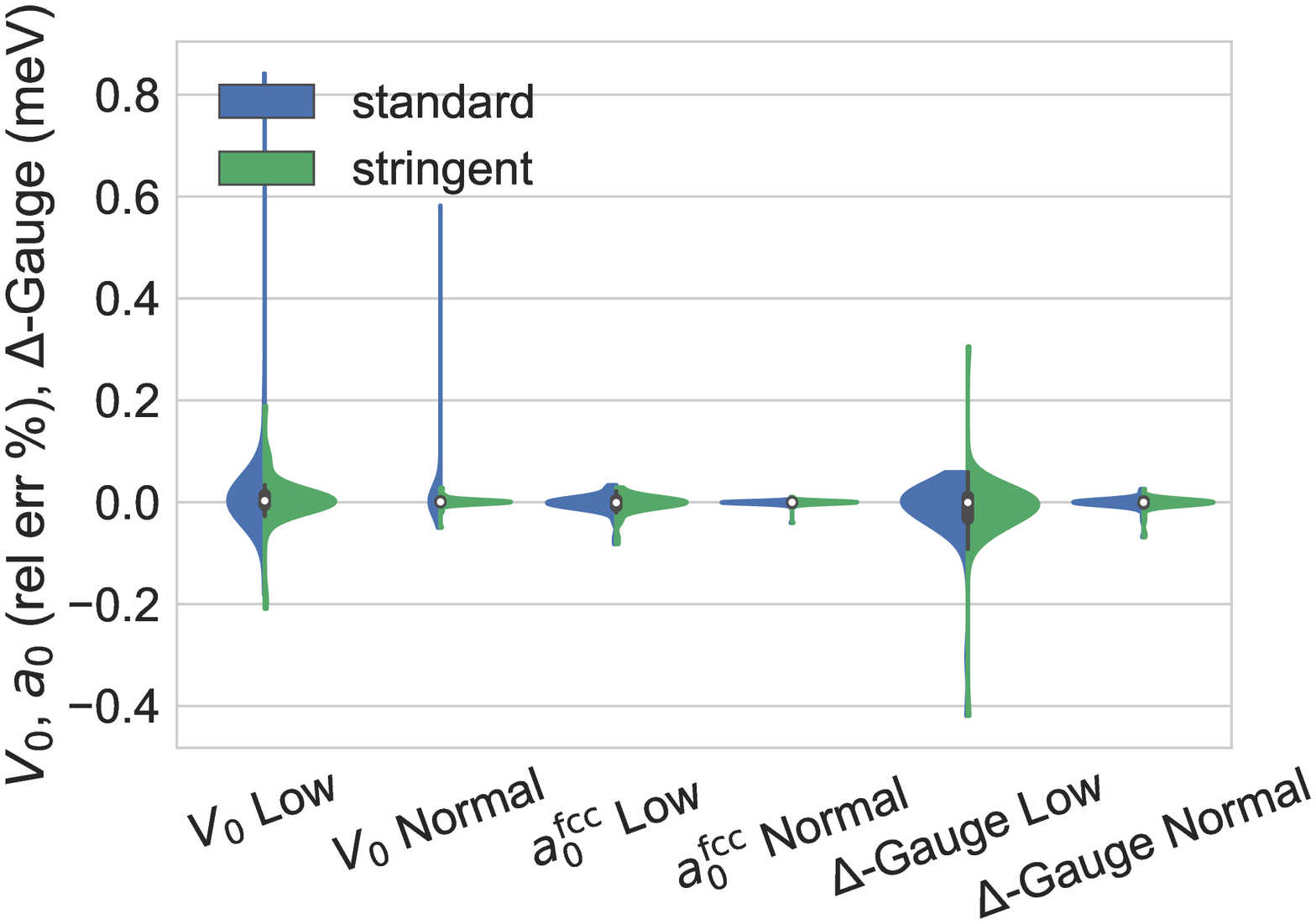}{Violin plot of the distribution of the error in $V_0$ and $a_0$ and \df\: at the low and normal hint as compared to their values at the high hint.}\selectlanguage{english}

Figure~\ref{violin_low_normals_errors_B} shows that the convergence of $B$ and $B_1$ is a factor of 10 to 100 slower than that of $V_0$ and the \df\: itself. 

\makefigure{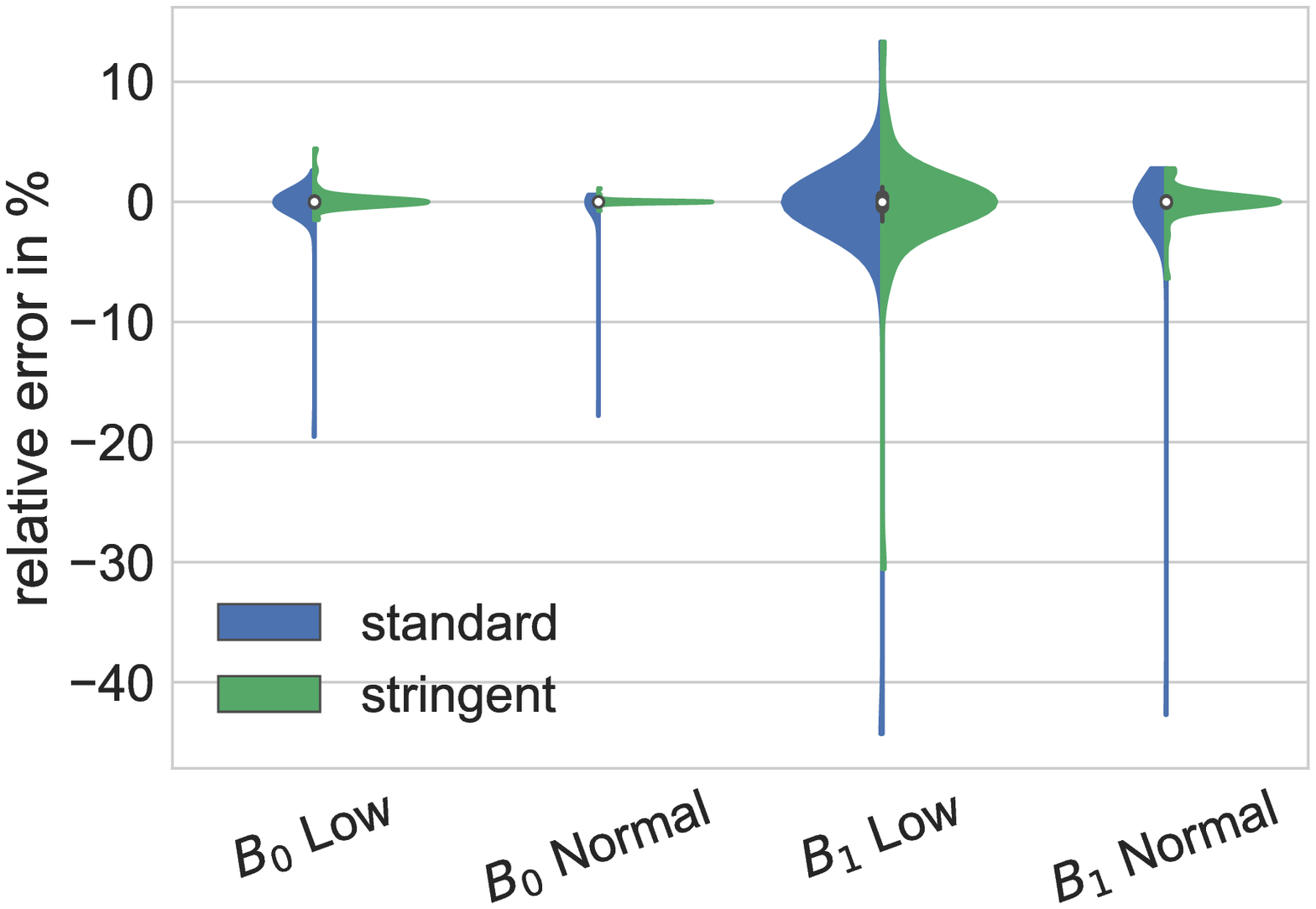}{Violin plot of the error in B and B1 at the low and normal hint as compared to their values at the high hint.}\selectlanguage{english}

\subsection{The GBRV dataset}

Complementary to the \df\: test we also perform the GBRV test on the two tables.~\cite{Garrity_2014} The GBRV tests consist of two parts. In the first test, the optimal lattice parameter of FCC and BCC single element structures are compared to AE reference values. In the second test, the lattice parameters of rocksalt, half-heusler and perovskite structures are compared. Reference values for noble gas FCC and BCC structures are however not present. We confirm the observation of Garrity \textrm{et al.} that the FCC and BCC results show a strong correlation, see section~\ref{sect:corr}. We hence only discuss here the FCC results. The noble gases are not present in the GBRV tests since the FCC and BCC structures do not bind in GGA-PBE. 

\makefigure{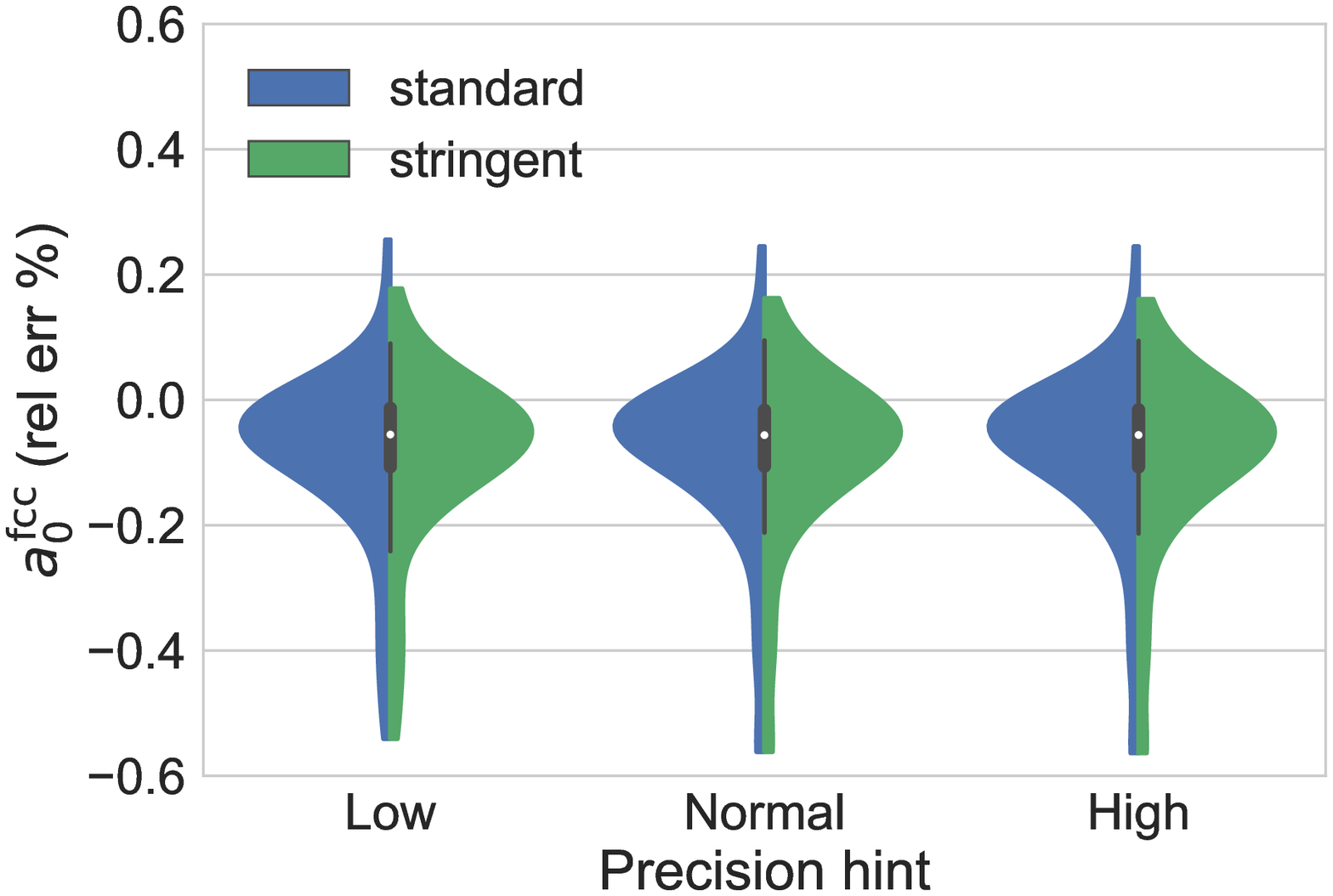}{Violin plot of the relative error on the lattice parameter of the GBRV test set.}\selectlanguage{english}

Figure~\ref{violin_gbrv_fcc} summarizes the distribution of the relative errors of the lattice parameter of the FCC GBRV test. Also for the GBRV test, we observe that both the low and normal hints already provide rather converged results, see also Fig.~\ref{violin_low_normals_errors}. In contrast to the \df\: tests, however, the high table does not significantly improve the GBRV test results. This difference partially relates to the fact that the GBRV tests do not contain any magnetic systems whereas the \df\:~tests do. In the \df\: tests we observe the strongest difference between the high and low tables for magnetic systems. Finally, we observe that in the GBRV tests the NCPPs tend to underestimate the lattice parameter with respect to the AE reference. The same trend is observed for the PAW data-sets that have performed the GBRV tests.~\cite{Garrity_2014,Jollet_2014} 

Besides the FCC and BCC elemental structures the GBRV reference data also contains 63 rocksalt structures, 54 half-heuslers (hH), and 138 perovskites (ABO3). The presence of these multiple-element systems allows for a real test of transferability. A full account of the GBRV compound test is given in the supplementary material. The high-table results do not differ significantly from the results obtained for the standard table. We therefore discuss here only the latter at normal $E_\mathrm{c}$ hint. Table~\ref{gbrv_errors} compares the performance of our standard table to that of various existing PAW data sets and USPP tables. Clearly all tables are of similar accuracy. Within the distribution, \ourtable\: does not perform best but it also does not perform worst in any of the structure types.  

\begin{table}\center\begin{tabular}{llllll}
\toprule
{} &  GBRV &  GBRV &  pslib &  VASP &  this \\
{} &  PAW &  USPP &  &  &   \\
\midrule
ABO3     &     0.089 &      0.078 &  0.200 & 0.127 & 0.185 \\
hH       &     0.126 &      0.111 &  0.144 & 0.140 & 0.116 \\
rocksalt &     0.129 &      0.121 &  0.216 & 0.150 & 0.184 \\
\bottomrule
\end{tabular}
\caption{GBRV average errors (relative error in \% in the lattice parameter) per compound group as compared to the GBRV-PAW, GBRV-USPP, pslib, and VASP results.~\cite{Garrity_2014}}\label{gbrv_errors}\end{table}\selectlanguage{english}

To investigate if the PSP for one specific element is performing badly in the GBRV compound test we summarize the results per element in Fig.~\ref{gbrv_elements}. We observe that our PSPs tend to slightly underestimate the AE lattice parameters although the distribution of our relative error is quite symmetric and peaked around the mean value. The other tables also tend to underestimate the AE reference but some with broader distribution. The elements that stand out most in the FCC and BCC tests F, S, and K also stand out in the compound test. Cs and Rb on the other hand perform better in the compounds than in the single element tests. 

\makefigure{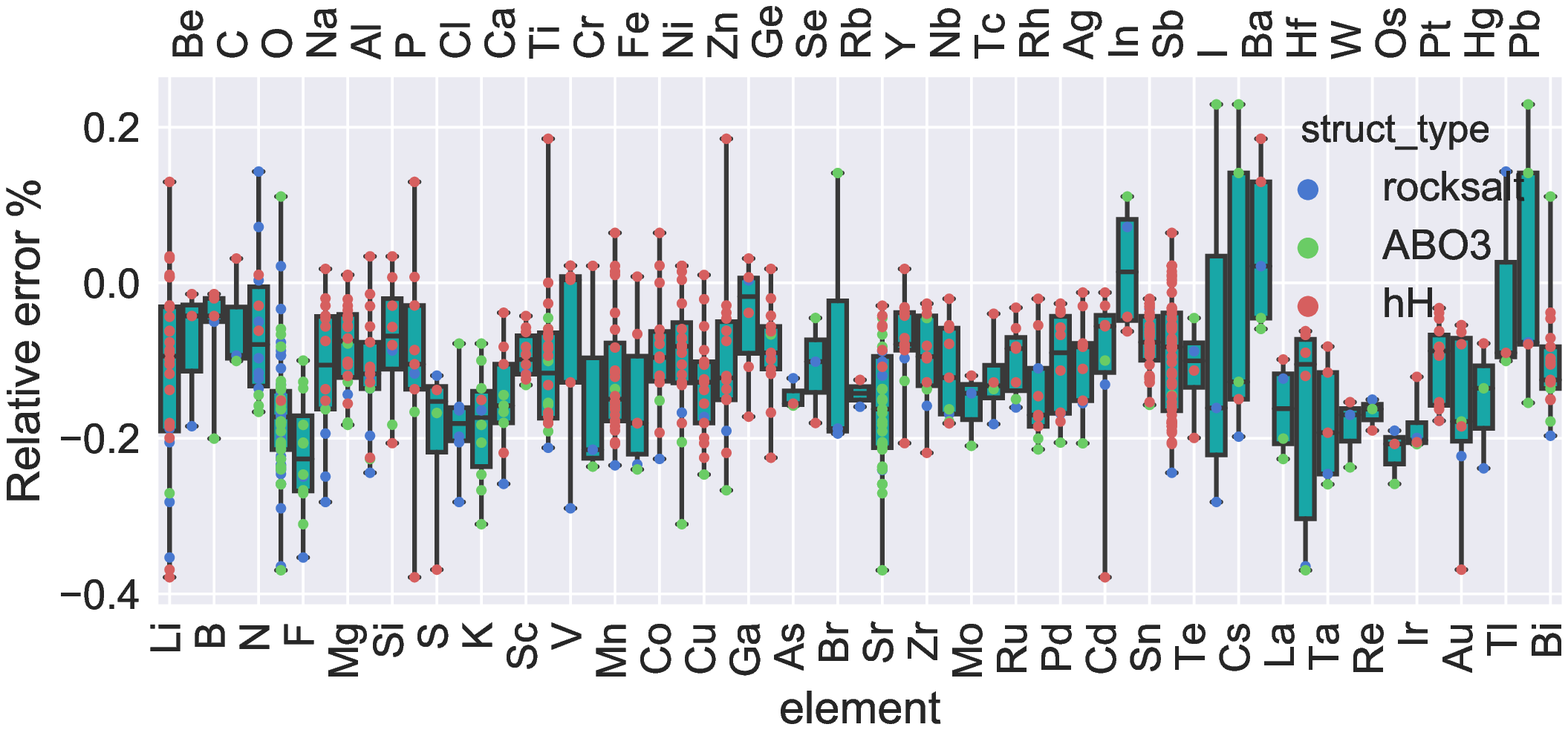}{GBRV compound tests error per element (figure to be put over two columns.)}\selectlanguage{english}

Finally, in Table~\ref{gbrv_compounds_outliers} we list those systems in the GBRV compound tests that have an error of more than 0.25\% with respect to the AE reference. 

\begin{table}\center\begin{small}
\begin{tabular}{llllllll}
\toprule
Formula &      Type &    AE &  GBRV &  GBRV &  pslib &  VASP &  this \\
&       &    &  PAW &  USPP &   &   &   \\
\midrule
  CdPLi &        hH & 5.969 &     5.957 &      5.955 &  5.945 & 5.952 & 5.946 \\
 SrHfO3 &      ABO3 & 4.155 &     4.141 &      4.148 &  4.133 & 4.146 & 4.140 \\
  LiAuS &        hH & 6.015 &     5.995 &      5.994 &  6.008 & 5.993 & 5.993 \\
    HfO &  rocksalt & 4.611 &     4.586 &      4.596 &  4.574 & 4.584 & 4.594 \\
    LiF &  rocksalt & 4.076 &     4.076 &      4.074 &  4.081 & 4.067 & 4.062 \\
  KNiF3 &      ABO3 & 4.039 &     4.035 &      4.036 &  4.037 & 4.036 & 4.026 \\
     VO &  rocksalt & 4.192 &     4.189 &      4.190 &  4.192 & 4.191 & 4.180 \\
   NaCl &  rocksalt & 5.714 &     5.702 &      5.701 &  5.696 & 5.701 & 5.698 \\
    LiI &  rocksalt & 6.038 &     6.023 &      6.020 &  6.030 & 6.021 & 6.021 \\
 SrLiF3 &      ABO3 & 3.884 &     3.883 &      3.881 &  3.884 & 3.884 & 3.873 \\
  KZnF3 &      ABO3 & 4.132 &     4.134 &      4.133 &  4.130 & 4.139 & 4.121 \\
 SrTaO3 &      ABO3 & 4.066 &     4.070 &      4.067 &  4.050 & 4.067 & 4.055 \\
 SrOsO3 &      ABO3 & 3.982 &     3.979 &      3.983 &  3.986 & 3.992 & 3.972 \\
    CaO &  rocksalt & 4.839 &     4.834 &      4.834 &  4.828 & 4.842 & 4.826 \\
\bottomrule
\end{tabular}
\end{small}
\caption{The outliers (relative error larger than 0.25 \%) in the GBRV compound tests. The columns list the lattice parameter of the conventional cell in Angstrom.}\label{gbrv_compounds_outliers}\end{table}\selectlanguage{english}

A general observation over all GBRV tests is that the elements that show larger deviations in our NCPP table like F also stand out in the PAW tables.~\cite{Garrity_2014,Jollet_2014} This seems to suggest that the error originates from the freezing of the core rather than from pseudizing the valence electrons. However for the most problematic elements, F, S, Cs, Rb, and K adding additional states to the valence partition turned out to be very difficult. 

\subsection{Ghost state detection}

The separable non-local operator that enters the pseudo Hamiltonian can lead to eigenstates for a given quantum number $l$ which are not ordered in energy by the number of nodes. As a result, eigenstates with nodes can appear with energies below the nodeless eigenstate, or the nodeless state can be followed directly by states with more than one node.~\cite{gonze90ghoststates,Gonze1991} The second projector in the \textsc{ONCVPSP} scheme is usually very efficient in removing these so-called ghost states in the occupied and low energy unoccupied energy range. The eventual appearance of ghost states at higher energies is, however, unavoidable. Since we aim at generating PSPs that can be used also to calculate properties requiring an accurate description of the unoccupied region, i.e. optical properties or $GW$ calculations, we explicitly test our PSPs for the presence of ghost states. This is done in the elemental solids by scanning the band structure and the density of states for dispersionless states, up to energies around 200~eV above the Fermi level. 
Ghost states could be removed in many cases by tuning the position of the second projector. Also, the addition of more semi-core states was found to improve the quality of the logarithmic derivative at high energies. Table~\ref{tab:ghosts} lists those PSPs that even after careful optimization of the input parameters still contain ghost states in the first 200~eV of unoccupied space and lists the `ghost-free' alternatives.

\begin{table}
\center
\begin{tabular}{lcl}
\toprule
Pseudo & $\epsilon$ & Alt Pseudo \\
\midrule
Mg   & 80 & Mg-sp\\
Cd   & 73 & Cd-sp\\
Sn-d & 60 & Sn-spd-high\\
Sb-d & 20 & Sb-spd-high\\
Te-d & 77 & Te-spd-high\\
Hg   & 66 & Hg-sp\\
Tl-d & 60 & Tl-spd-high\\
Pb-d & 70 & Pb-spd-high\\
Bi-d & 70 & Bi-spd-high\\
Po-d & 58 & Po-spd-high\\
\bottomrule
\end{tabular}
\caption{Pseudopotentials that have a ghost state present in the first 200 eV of unoccupied space. $\epsilon$ gives the energy (eV) at which the ghost appears. Alt Pseudo provides the high-accuracy alternative that does not show any sign of ghost states.}\selectlanguage{english} \label{tab:ghosts}
\end{table}

Note that the ghost states listed here are all so high in energy that for ground state calculations they do not cause any problem. Only for applications that require an accurate description of the unoccupied space as well (like $GW$ and optical properties), do the nonphysical resonances introduced by the ghost states lead to incorrect results. 

\subsection[]{Phonon modes at $\Gamma$}
\label{sect:phonon}

Calculating the phonon modes at $\Gamma$ allows for the evaluation of two useful quantities even when no reference values are available. First, it allows for an evaluation of the rate of convergence of the optical modes. Second, evaluating how strongly the acoustic sum rule is broken for the acoustic modes provides another test for the PSPs.

\makefigure{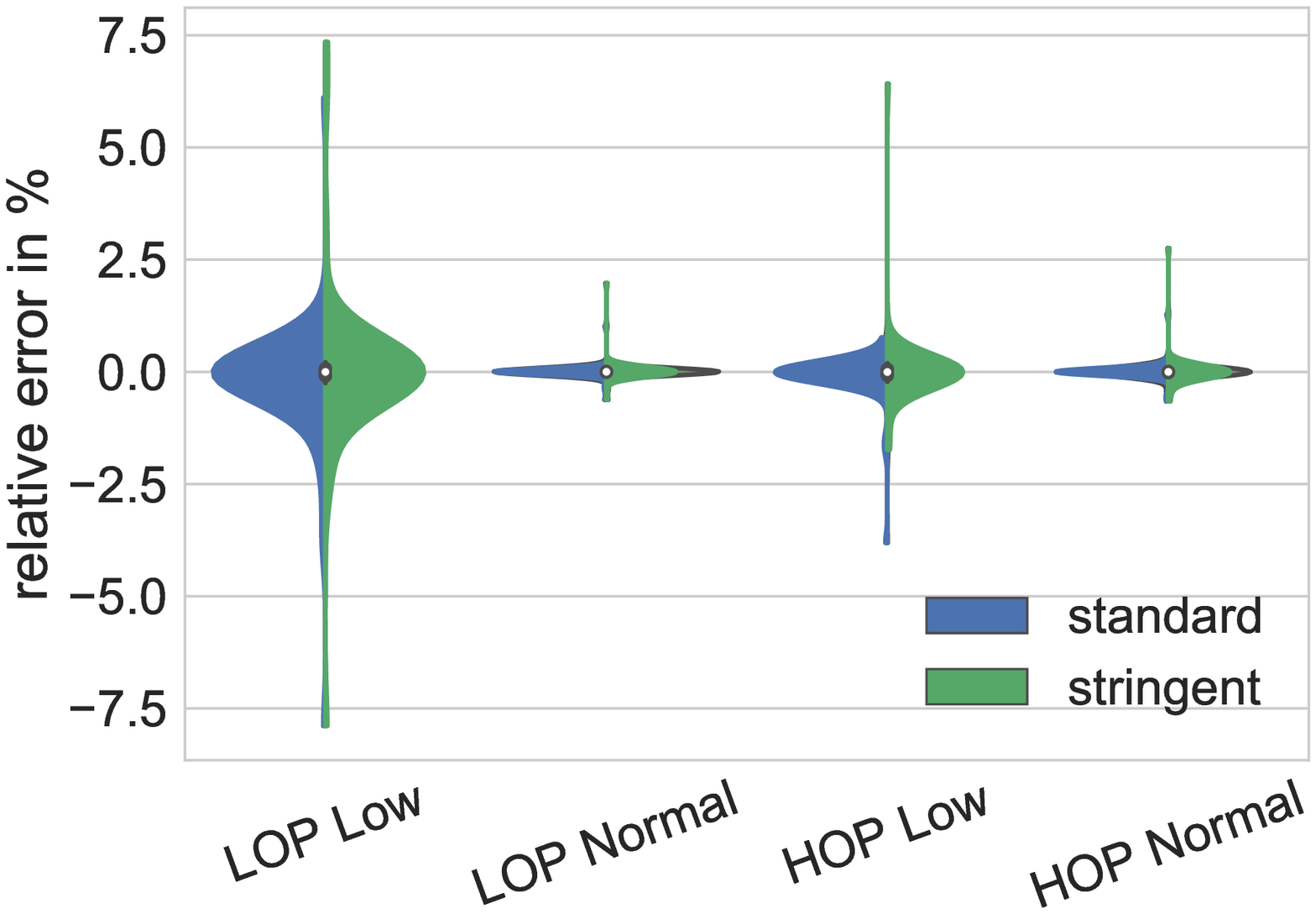}{Violin plot of the distribution of the relative errors in the highest optical phonon (HOP) and lowest optical phonon (LOP) at $\Gamma$ at the low and normal hint as compared to their values at the high hint, after enforcing the acoustic sum rule.}\selectlanguage{english}

The convergence of the optical phonon modes is illustrated in Fig.~\ref{violin_low_normals_errors_phonons_asr}. In contrast with properties like the equilibrium volume and the \df\: the phonon modes are by far not converged at the low $E_\mathrm{c}$ hint. 

\makefigure{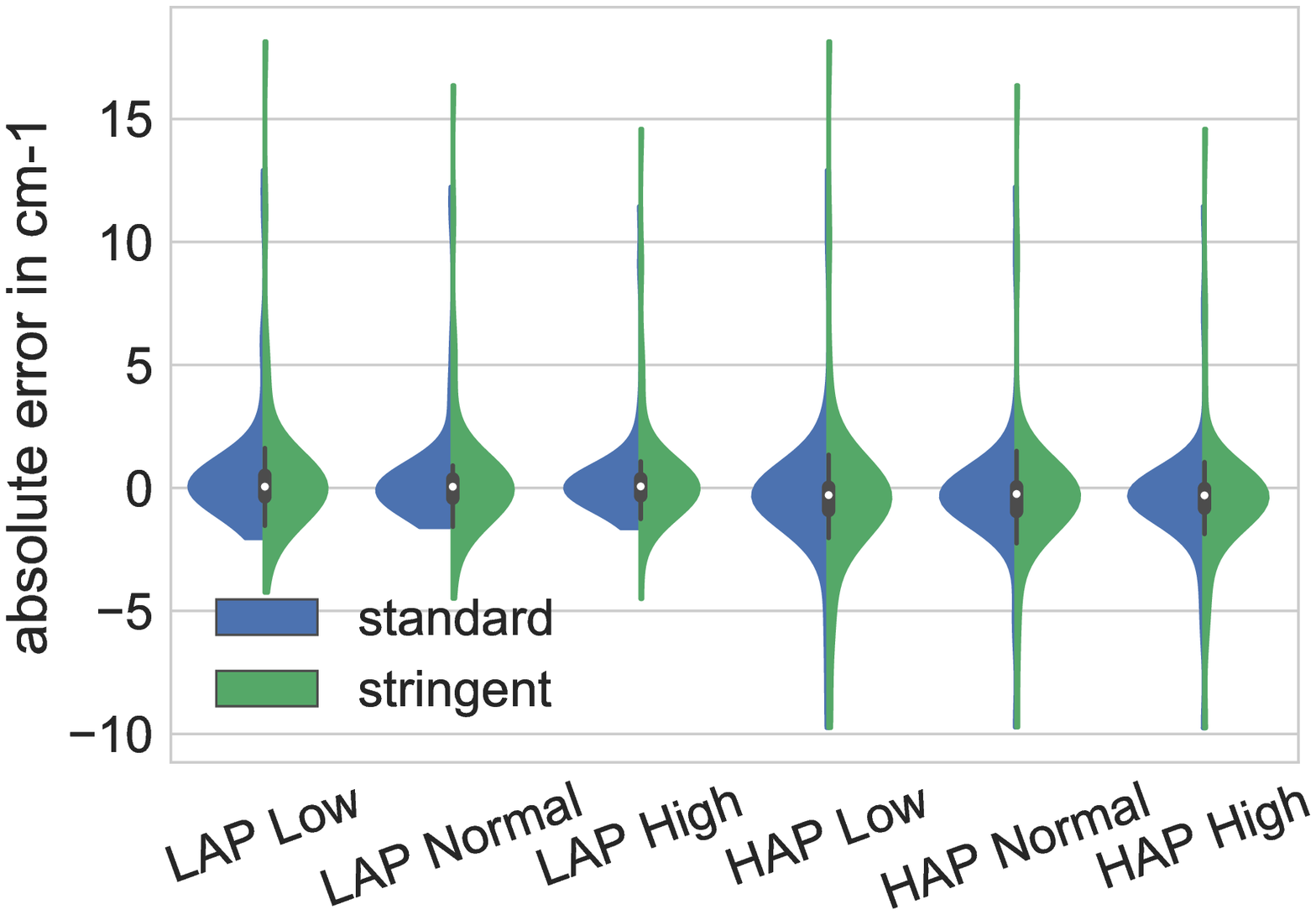}{Violin plot of the distribution of the absolute errors in the highest acoustic phonon (HAP) and lowest acoustic phonon (LAP) at $\Gamma$ not enforcing the acoustic sum rule. The outliers are Ne ($\sim$14), Mg ($\sim$11), Na ($\sim$6), Ge ($\sim$4), an Cu ($\sim$4) all in cm$^{-1}$.}\selectlanguage{english}

The breaking of the acoustic sum rule (ASR) is shown in Fig.~\ref{violin_errors_phonons_noasr}, except for a few outliers the error remains within 2 cm$^{-1}$. We note that, although only slightly, the error is larger in the high table than in the standard table. This is caused by the harder pseudopotentials. All of the values obtained are easily corrected by standard techniques for imposing the ASR.

\subsection{Correlations between the tests}
\label{sect:corr}

Performing different tests makes sense provided the results do not correlate strongly. To investigate the correlation between the different tests, the correlation matrix between the GBRV FCC and BCC, \df, \dfp, and the absolute error in the acoustic sum rule for the phonon mode is shown in figure~\ref{all_cor_mat}.  

\makefigure{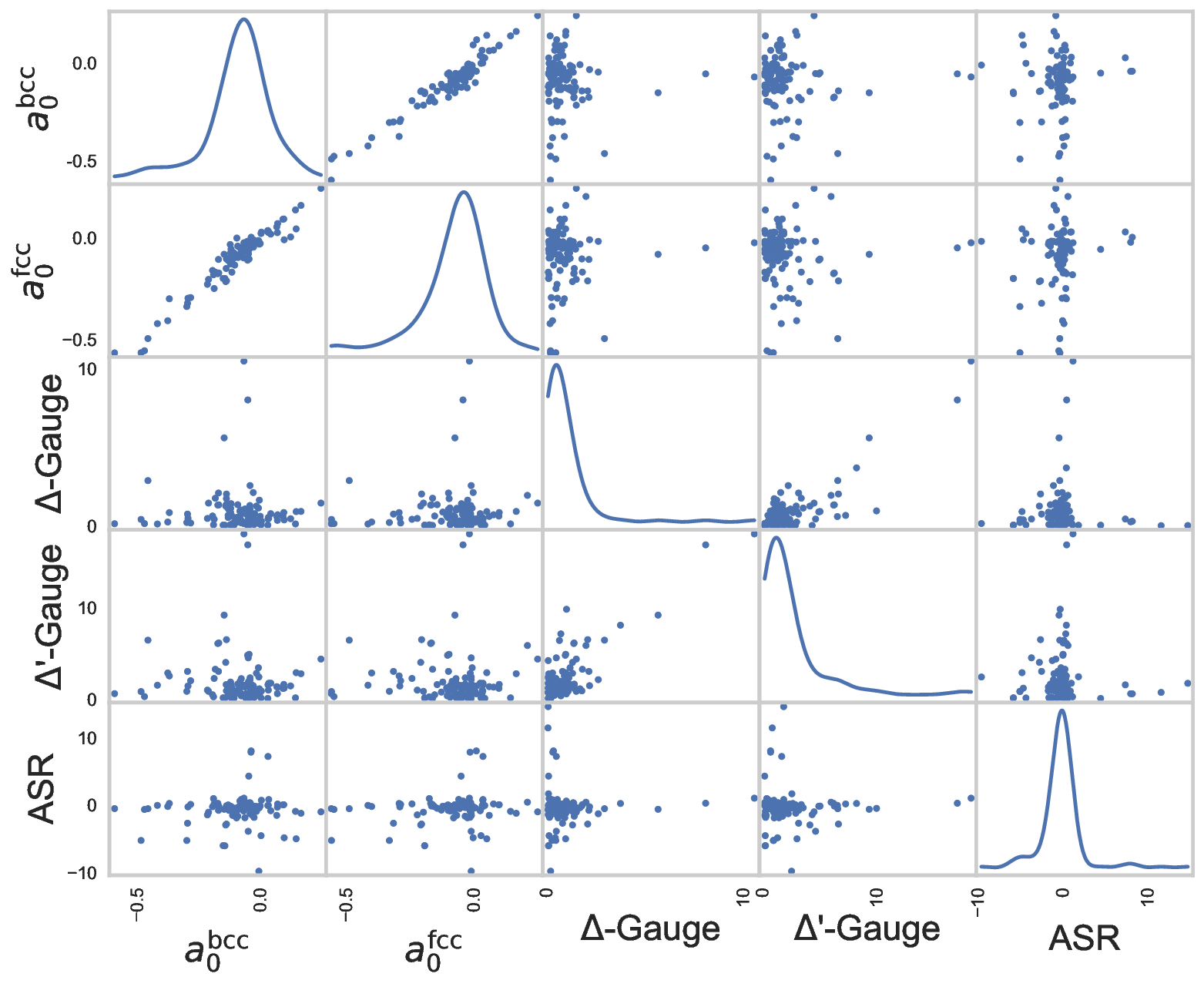}{Correlation between the results of the tests in the \textsc{PseudoDojo} including all PBE pseudopotentials. FCC and BCC denote the relative errors in the lattice parameters of the GBRV tests, Delta and Delta prime the \df\: and \dfp\: tests and ASR the absolute error in the first acoustic mode. The diagonal shows the distribution of the test results.}\selectlanguage{english}

As indicated above, the FCC and BCC GBRV tests show a very strong correlation which means that performing both does not add additional information. \df\: and \dfp\: also show some correlation, as expected, but considering both still adds information. The GBRV tests and the two \df\: test on the other hand hardly show any correlation. The error on the acoustic phonon modes finally seems to be completely decoupled from all other tests.

\section{Discussion of individual pseudopotentials}
\label{sect:discussion}

\subsection{H, He}  

The $1s$ wavefunctions in H and He are rather localized. One should therefore exercise special care to find values for the pseudization radii that give a good compromise between accuracy and efficiency. In H the pseudization radius for the $1s$ is set to 1.0~a.u and 1.25~a.u in He. Both PSPs contain two $s$ projectors. 

\makefigure{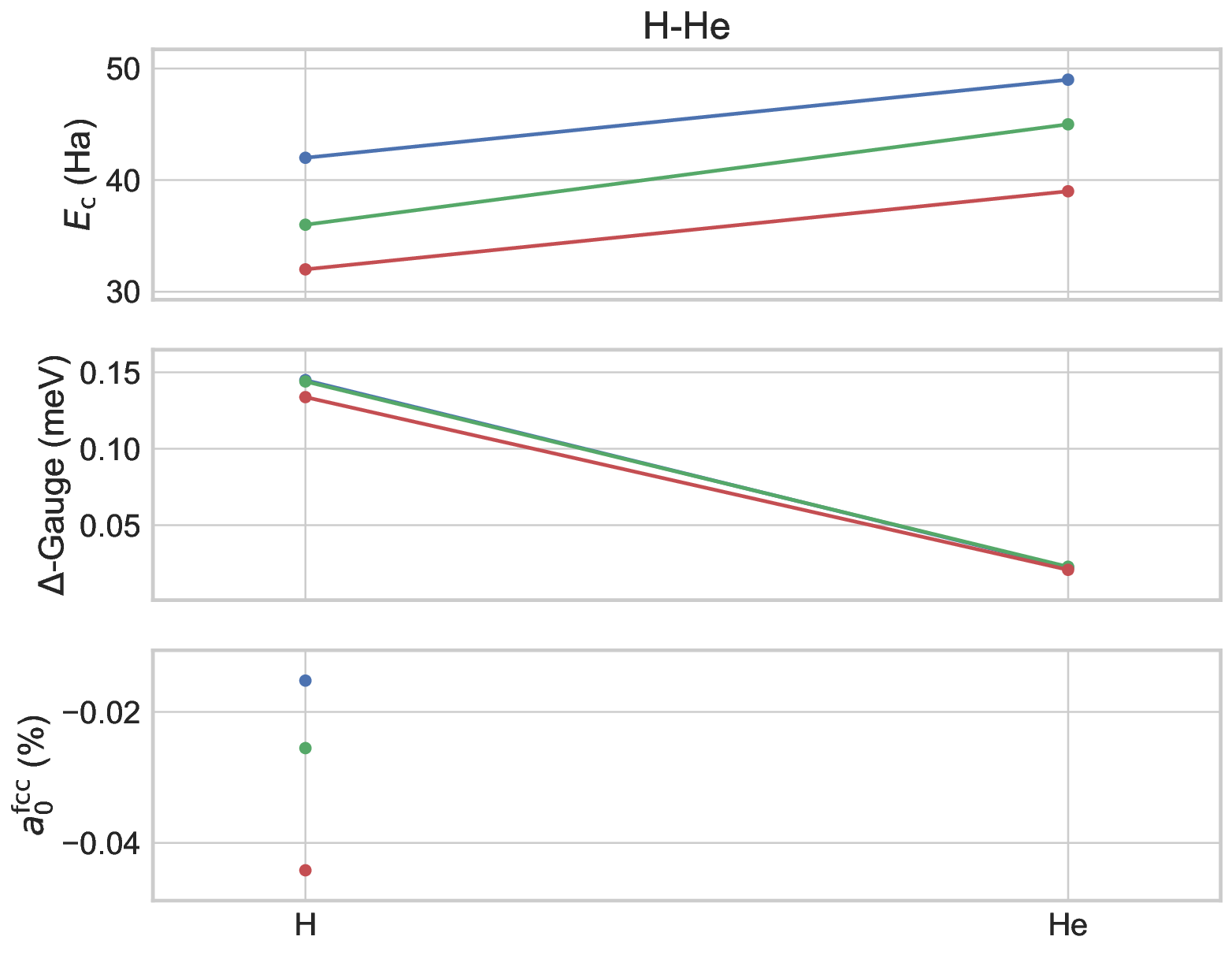}{The main test results for H and He PSPs. The blue, green, and red data are calculated at the high, normal, and low $E_\mathrm{c}$ hints respectively. The noble gases are not present in the GBRV tests since the FCC and BCC structures do not bind in GGA-PBE.}

The $p$ orbitals in H and He are not bound in GGA-PBE hence we use only one projector for the $p$ channel.
The main test results for the H and He pseudopotentials are shown in Fig.~\ref{H-He}.

\subsection{Li, Be}  

In Li and Be, the $1s$ states are more localized than in H and He and the $p$ orbitals are bound. We include the $1s$ electrons in the semicore, which yield PSPs that are more transferable and accurate, at the price of a non-negligible increase in the $E_\mathrm{c}$, see Fig.~\ref{Li-Be}. For this reason, the \ourtable\: provides two versions for elements. The standard version uses a $s$ channel cutoff radius of 1.4~a.u for Li and 1.35~a.u for Be  with an indicative $E_\mathrm{c}$ of 35~Ha and 42~Ha, respectively. The local part of the PSPs is obtained by pseudizing the AE potential and two projectors both for $s$ and $p$. The high accuracy versions mainly differ from the standard ones in the use of smaller $r_c$ for the $s$ channel (1.2 a.u. both for Li and Be) and, consequently, have a slower convergence in reciprocal space. 

\makefigure{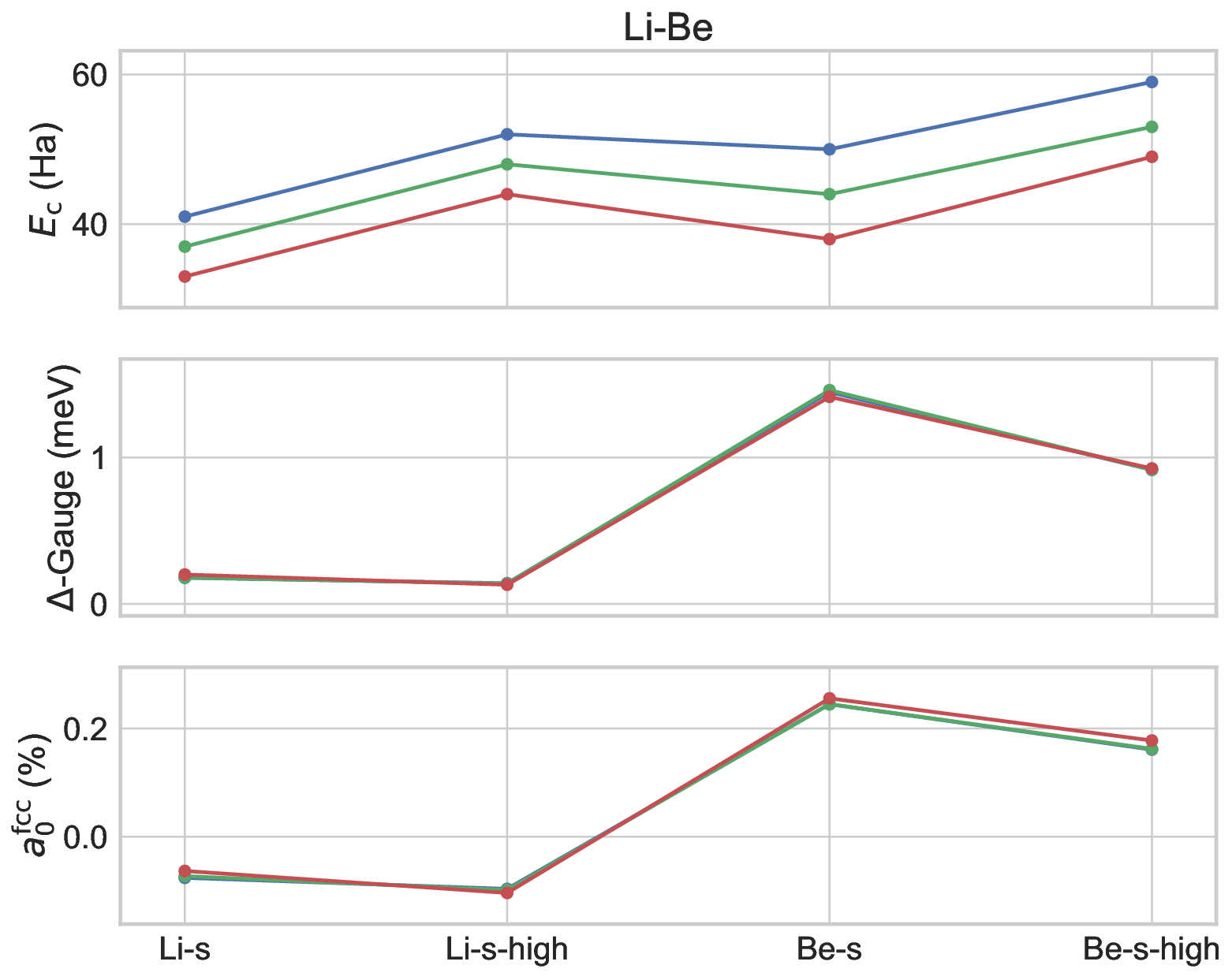}{The main test results for Li and Be PSPs. The blue, green, and red data are calculated at the high, normal, and low $E_\mathrm{c}$ hints respectively.}

\subsection{B, C, N, O, F, Ne} 

In this set of elements, the $1s$ states are in the core, and the $E_\mathrm{c}$ is governed by the spatial localization of the $2p$ states. The choice of $r_c$ for the $p$ channel has therefore a significant impact on the transferability of the PSPs. We use two projectors per angular channel and a pseudized version of the AE potential as local part. The maximum angular momentum explicitly included in these PSPs is the $p$ channel, $l_{max}=1$. For O and F in addition a single $d$ projector is added to improve transferability. An overview of the evolution of the main test results for these PSPs is shown in Fig.~\ref{B-Ne}.

F is one of the elements for which the GBRV tests show the largest error. This is also observed in the PAW tables that have been tested with the GBRV test. In addition it is observed that the F PSPs perform badly in describing atomization energies or molecular systems.~\cite{goedecker2017}

Ne is one of the few elements with frozen core states for which adding non-linear core corrections does not improve transferability. 
The AE core is rather localized and therefore difficult to model without spoiling convergence.
Especially the equation of state curves obtained in the in the \df\: calculations tend to be far from the reference curves. Moreover, solving the electronic self-consistency problem turned out to be unstable for many of the model core charges tried. 

\makefigure{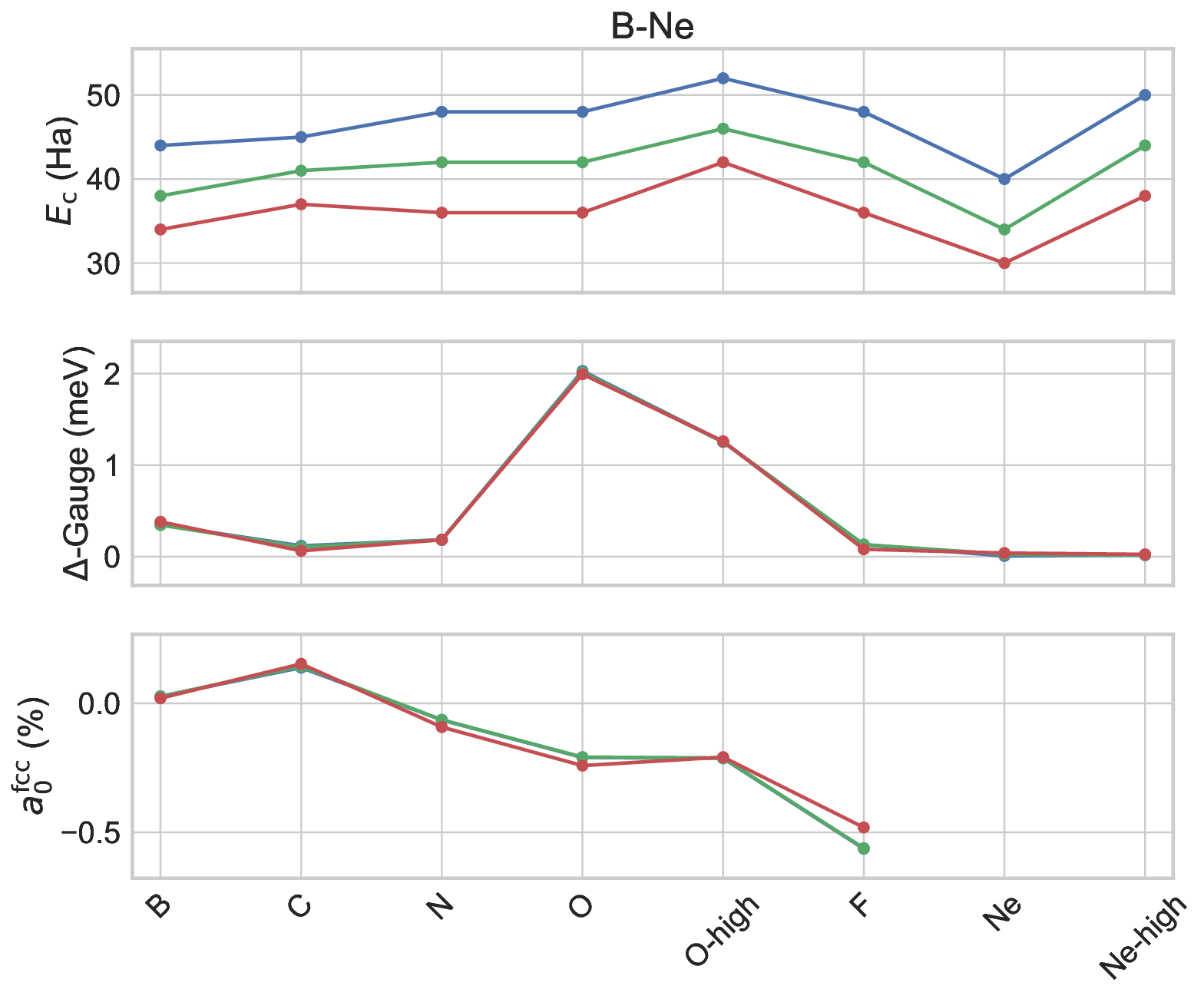}{The main test results for B to Ne PSPs. The blue, green, and red data are calculated at the high, normal, and low $E_\mathrm{c}$ hints respectively.}

\subsection{Na, Mg} 

In both PSPs with $2s$ and $2p$ in the valence no non-linear core corrections are applied. As for Ne, the very strong localization of the 1s core makes creating a transferable model core charge very complicated. Adding the $2s$ and $2p$ significantly improves both the \df\: and GBRV test results. In addition, for Mg, the ghost state at around 80~eV is removed.

\makefigure{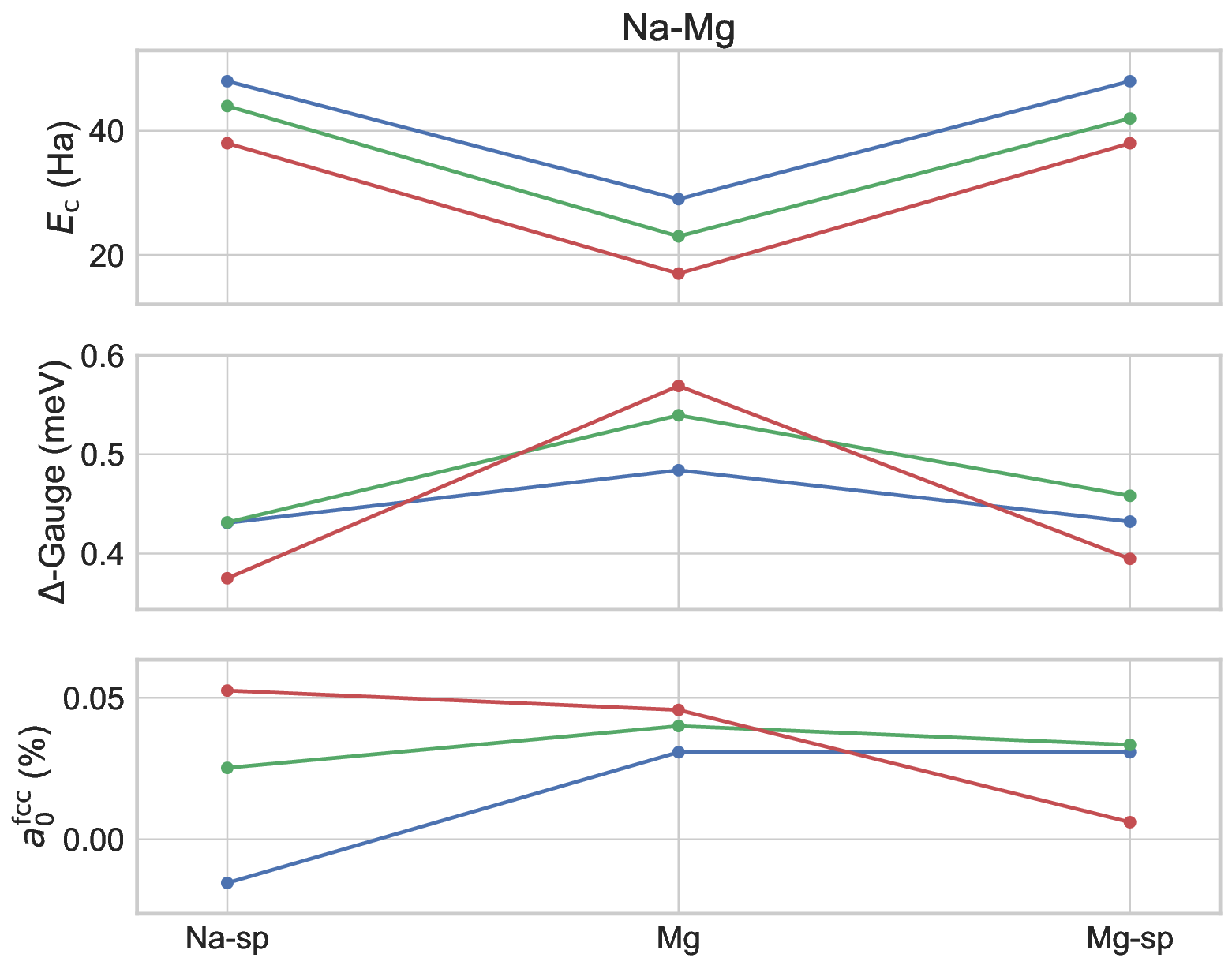}{The main test results for Na and Mg PSPs. The blue, green, and red data are calculated at the high, normal, and low $E_\mathrm{c}$ hints respectively.}

\subsection{Al, Si, P, S, Cl, Ar}  

In this series the $2s$, $2p$ and $3s$ states are full and the $3p$ orbitals are gradually filled. The shell with $n=2$ is well separated from the $n=3$ electrons and can be safely frozen in the core. Moreover the $3s$ and $3p$ electrons are rather delocalized and their pseudization does not pose any problem in the NC formalism. For these elements, it is common practice to include $d$ projectors in order to improve the transferability.  

For the purpose of convergence studies and the comparison to AE results we also provide a version with $2s$ and $2p$ in the valence for this series of elements. The high $E_\mathrm{c}$ required for these PSPs and non-systematic accuracy improvement make them however hardly useful for standard application. They are therefore not part of the high table. 

\makefigure{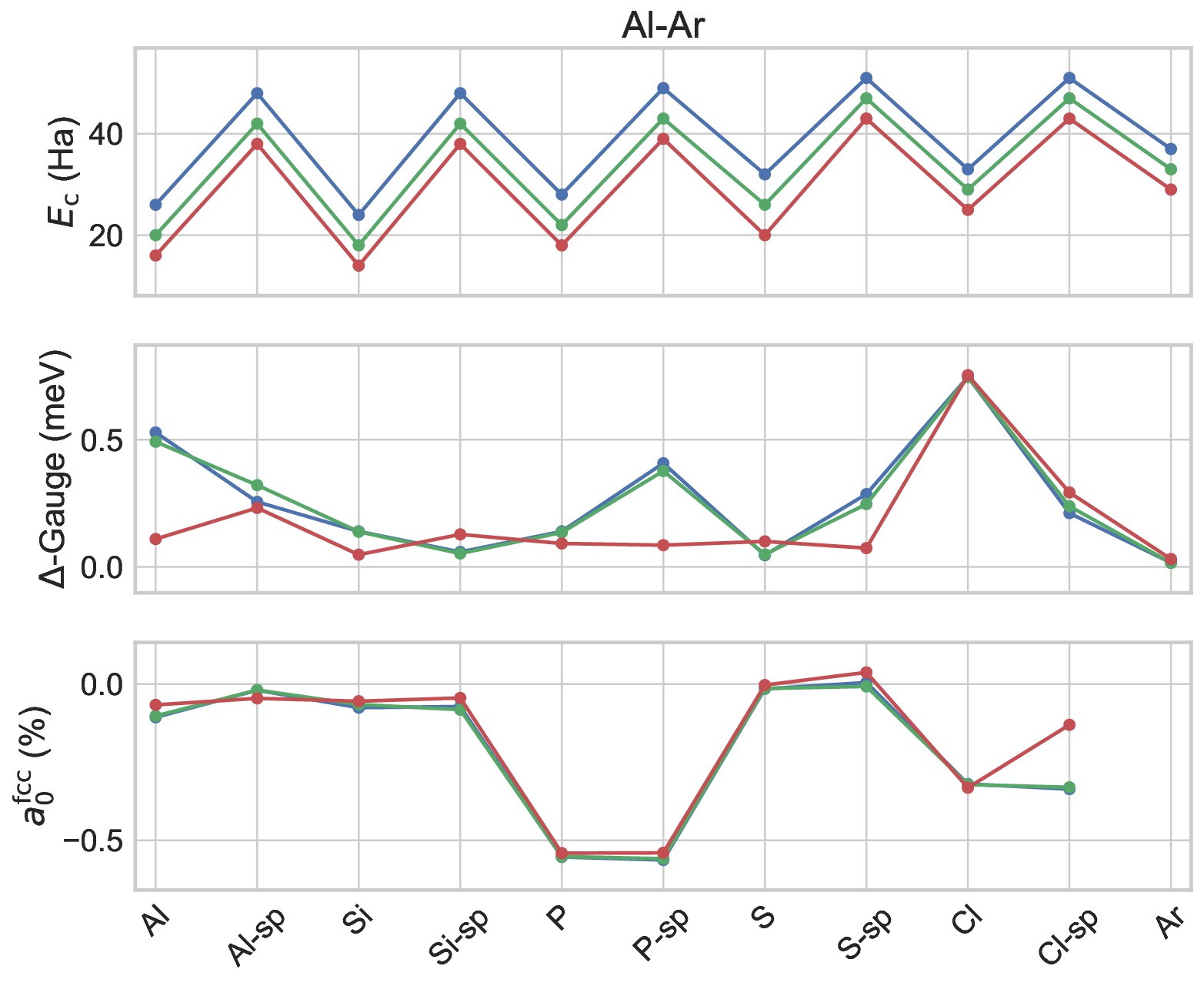}{The main test results for the Al to Ar PSPs. The blue, green, and red data are calculated at the high, normal, and low $E_\mathrm{c}$ hints respectively.}

\subsection{K, Ca} 

The default versions for these two elements have the $3s$ and $3p$ in the valence and contain two $d$ projectors to improve transferability. Given the reasonable $E_\mathrm{c}$ hints and the good test results, see Fig.~\ref{K-Ca}, these are part both of the standard and high table. 

\makefigure{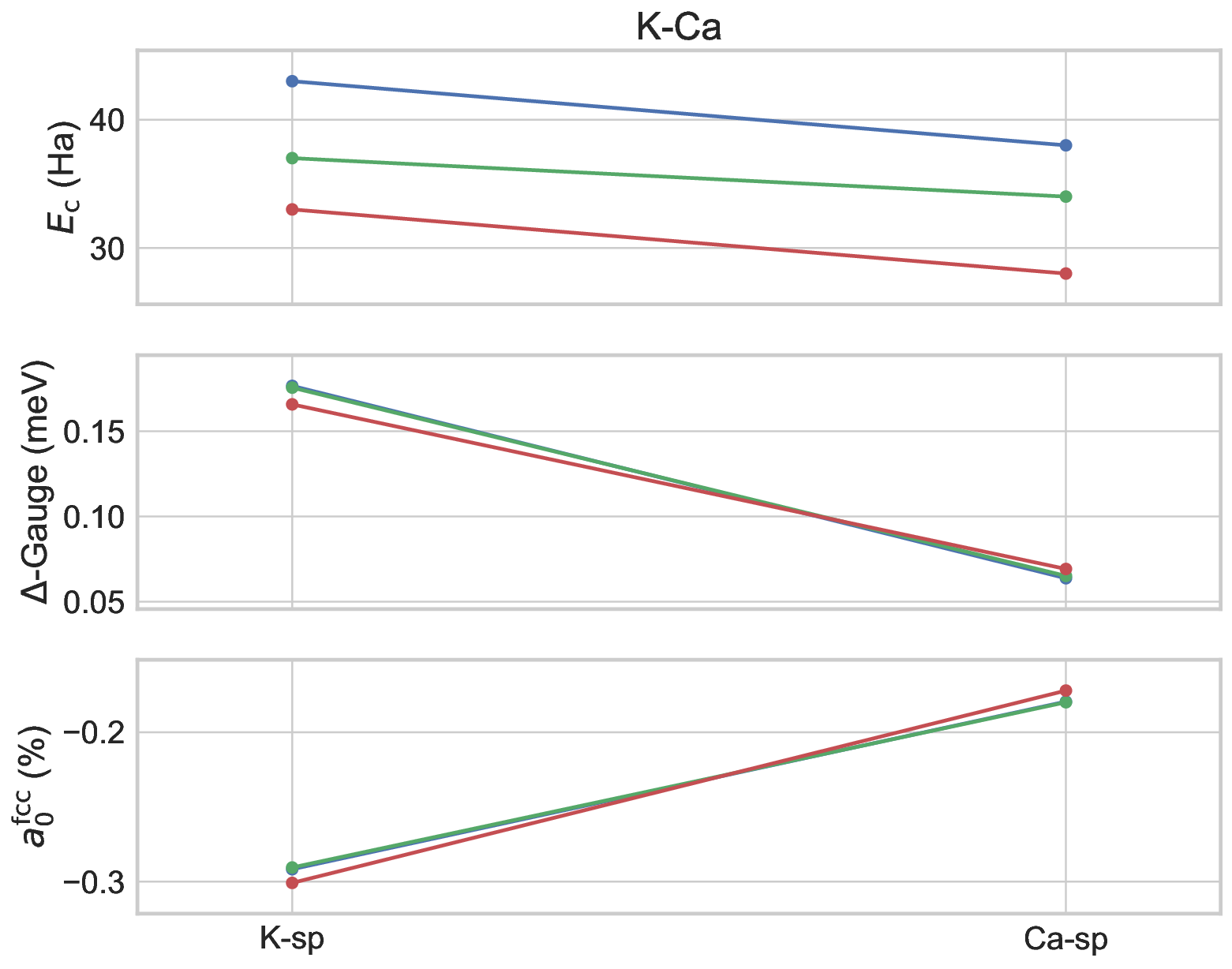}{The main test results for the K and Ca PSPs. The blue, green, and red data are calculated at the high, normal, and low $E_\mathrm{c}$ hints respectively.}

\subsection{3d transition metals}   

For the $3d$ transition metals, the $3s$ and $3p$ states are part of the valence partition. For both Fe PSPs, the degree of continuity at the pseudization radius was lowered to the third derivative. Generation a PSP for Fe with continuous derivatives up to fourth order at the pseudization radius leads to prohibitively large requirements on the $E_\mathrm{c}$.

The most complicated elements in this series are Cr, Mn, and Fe. Especially obtaining a PSP that performs well in the magnetic structures of the \df\: test is very hard. The standard versions, with a still reasonable $E_\mathrm{c}$ hint, have \df\: results that are well beyond what is usually considered acceptable (see Fig.~\ref{3d-transition-metals}). The high accuracy version fixes this, however at the cost of a considerable increase in the $E_\mathrm{c}$ needed. Both the standard and the high versions, however perform equally well in the (non-magnetic) GBRV tests. 

\makefigure{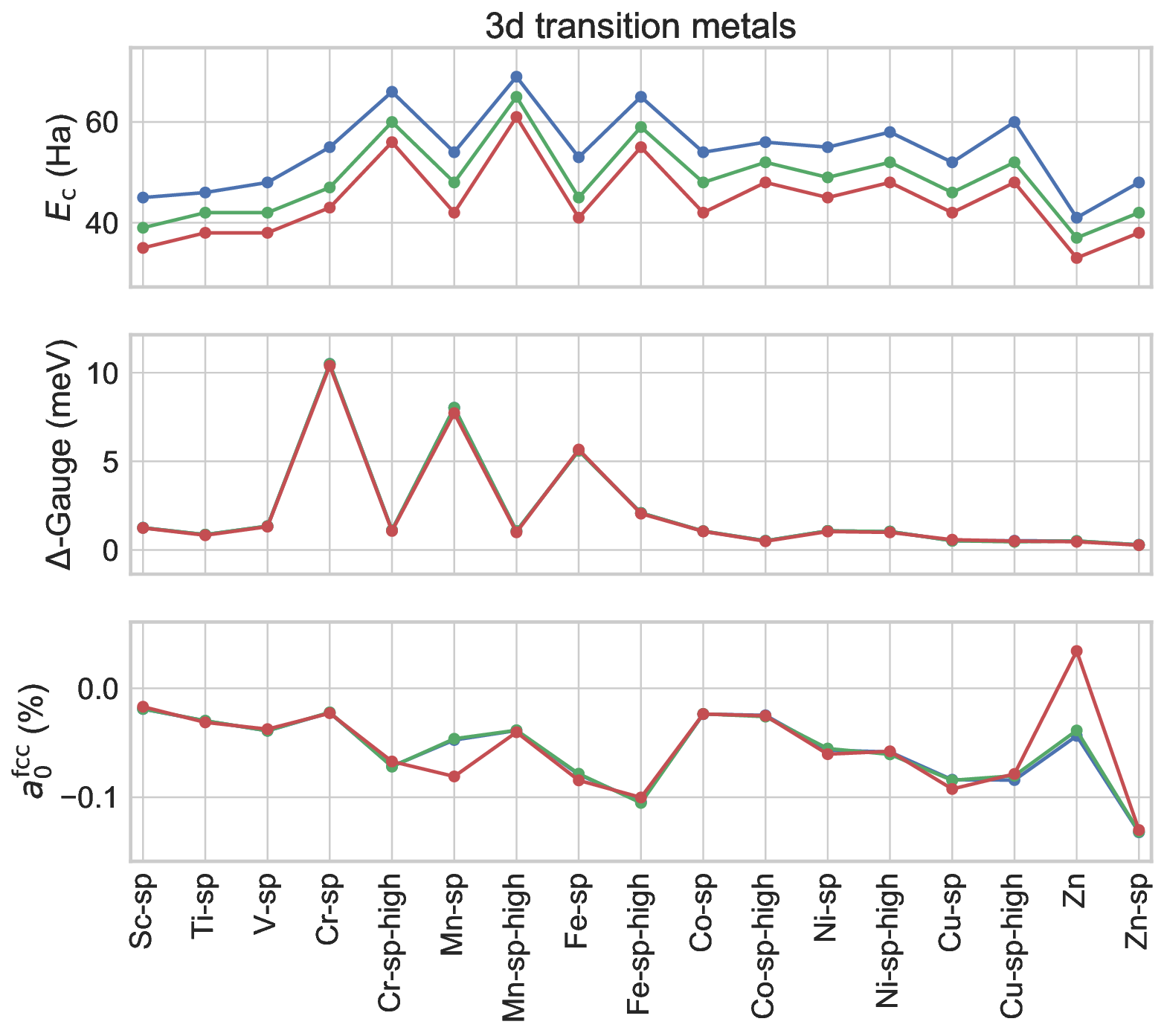}{The main test results for the $3d$ transition metal PSPs. The blue, green, and red data are calculated at the high, normal, and low $E_\mathrm{c}$ hints respectively.}

\subsection{Ga, Ge, As, Se, Br, Kr} 

In these elements, the $3d$ shell is full and we have a progressive filling of the $3p$ states. The $3d$ electrons, however, overlap with the $3p$ states and therefore can play a role in determining the physical properties of a crystalline system. For this reason, our standard table contains pseudopotentials with $3d$ electrons in valence for Ga, Ge, As, Se, while $3d$ electrons are frozen in Br and Kr. This is our recommended configuration albeit the presence of the localized $3d$ states leads to a relatively large $E_\mathrm{c}$, see Fig.~\ref{Ga-Kr}. A version of Ga-Ge-As-Se with the $3d$ electrons frozen in the core is also available for low $E_\mathrm{c}$ applications. 

For Br (and as well for I) we also provide a version with $3s$, $3p$, and $3d$ ($4s$, $4p$, $4d$ for Iodine) in the valence. These are provided mainly for the use in accurate $GW$ calculations in which the inclusion of entire electronic shells can be important, see e.g. Ref~\citenum{scherpelz16} for the example on I. For ground state calculations the $3d$ ($4d$) valence has been found to be sufficiently accurate and is therefore the choice for the stringent table. 

\makefigure{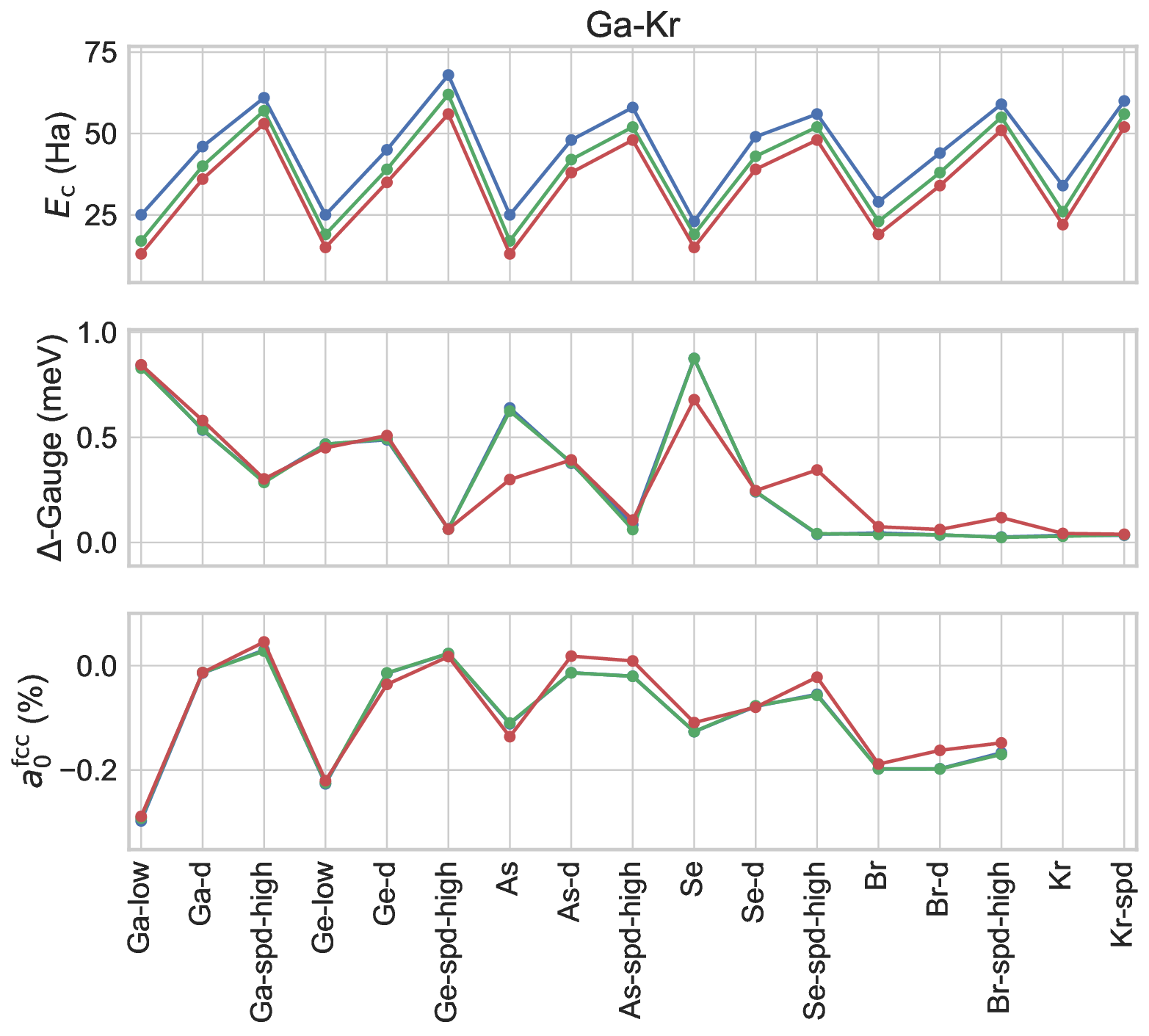}{The main test results for the Ga to Kr PSPs. The blue, green, and red data are calculated at the high, normal, and low $E_\mathrm{c}$ hints respectively.}

\subsection{Rb, Sr} 

The main test results and $E_\mathrm{c}$ hints for the PSPs for Rb and Sr are shown in Fig.~\ref{Rb-Sr}. For both elements, very reasonable $E_\mathrm{c}$ hints can be achieved. The alkaline elements from Rb downward start to show a decreasing performance in the GBRV test. As for F and S, this is in line with the results obtained for PAW data sets in the GBRV tests. Attempts to make harder, more accurate PSPs did not lead to improved GBRV results. 

\makefigure{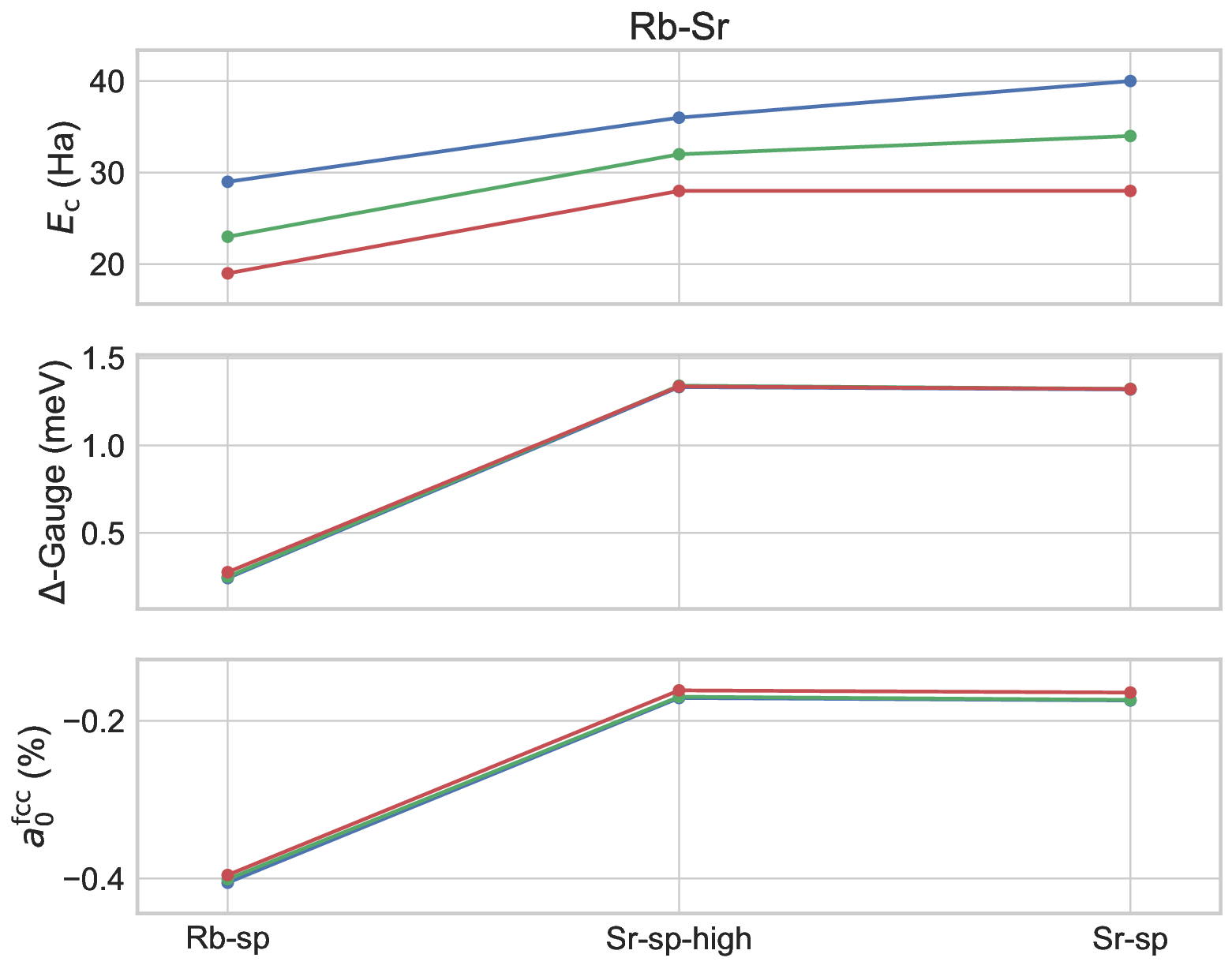}{The main test results for the Rb and Sr PSPs. The blue, green, and red data are calculated at the high, normal, and low $E_\mathrm{c}$ hints respectively.}

\subsection{4d transition metals} 

The PSPs for the $4d$ transition metals all contain the $4s$ and $4p$ states in the valence. This leads to both reasonable $E_\mathrm{c}$ energies and test results, see figure~\ref{4d-transition-metals}. Only Ru and Rh have \df\: results that are only barely acceptable. The \dfp\: results for these two PSPs (1.5 and 2.2) are still well within the acceptable range. The relatively high bulk modulus of these two elemental solids (310 and 250) causes the high \df\: values. For Cd, finally, we provide three versions. 
The version with the $4s$ and $4p$ states in the core (Cd) gives the best results for the \df\: 
but the GBRV is far from ideal. 
Including the $4s$ and $4p$ states in the valence (Cd-sp) improves the GBRV results at the price of a non-negligible increase of $E_\mathrm{c}$ while the \df\: worsens. 
Decreasing the core radius in the high-accuracy version (Cd-sp-high) leads to acceptable GBRV and \df\: results but at the cost of a larger $E_\mathrm{c}$.
Our standard table includes Cd-sp while the stringent table uses Cd-sp-high. 

\makefigure{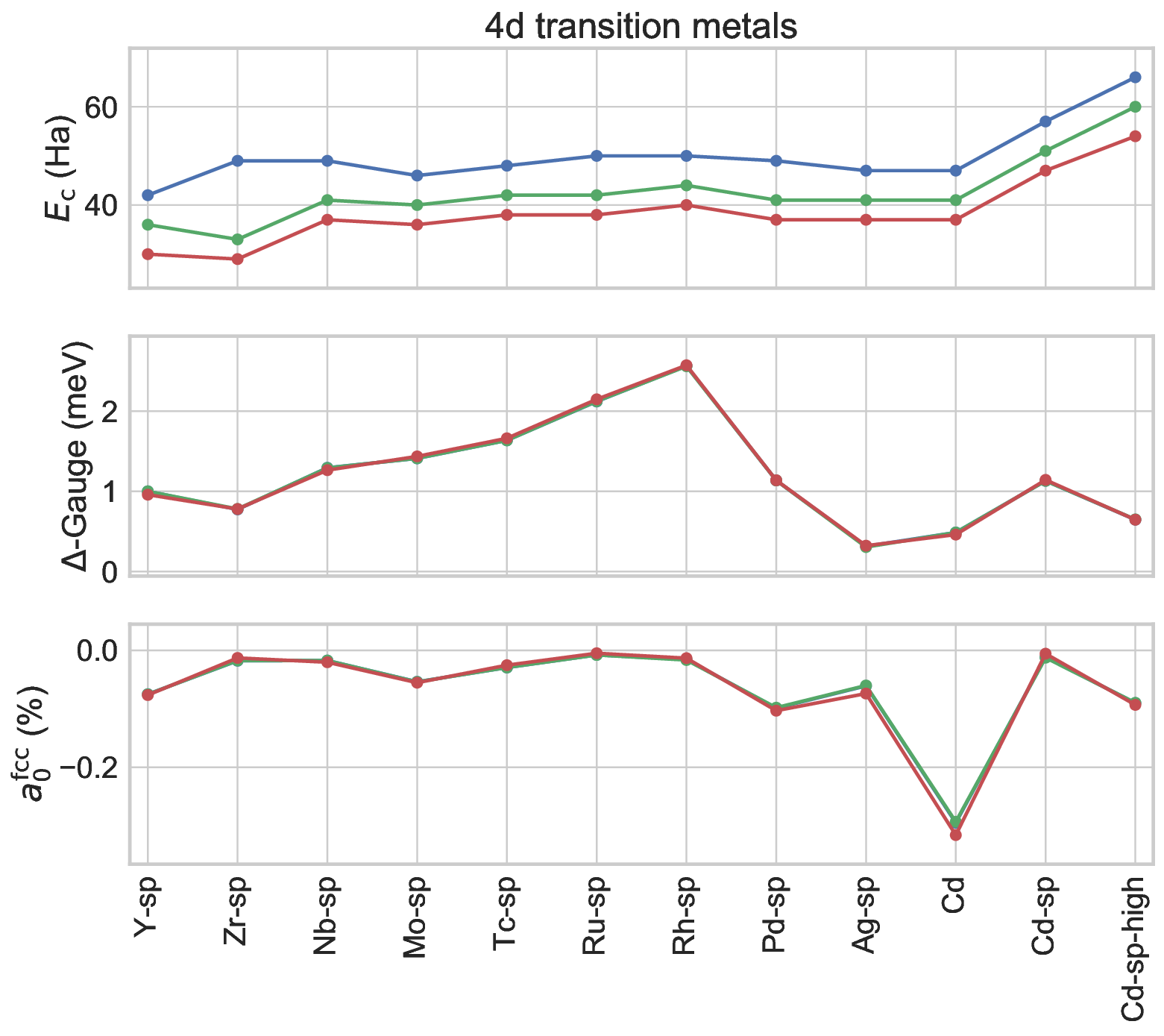}{The main test results for the $4d$ transition metal PSPs. The blue, green, and red data are calculated at the high, normal, and low $E_\mathrm{c}$ hints respectively.}

\subsection{In, Sn, Sb, Te, I, Xe}  

In the series from In to I, we provide three different versions of core-valence partitioning: no $n=4$ states in the valence (except for In), $4d$ in the valence, and the full $n=4$ shell in the valence. For all these pseudopotentials \df\:, \dfp\: and GBRV are well within the acceptable range. The main difference lies in the description of the unoccupied space. The PSPs for which all $n=4$ states are frozen in the core show deviations in the logarithmic derivative starting around 3 Ha above the Fermi level. Including the $4d$, which lie 0.7-1.5~Ha below the Fermi level, introduces ghost states in the elemental solid between 20 and 80~eV above the Fermi level. Finally including the full $n=4$ shell we see no sign of ghost states and the logarithmic derivatives agree well up to 7-10~Ha above the Fermi level. The cost for this accuracy is an increase in $E_\mathrm{c}$ of 20-30~Ha, see Fig.~\ref{In-Xe}. 

For Xe we freeze the full $n=4$ shell. The tests did not reveal any ghost states but the logarithmic derivative shows a sharp deviation around 4~Ha above the Fermi level.
A version with the full $n=4$ shell in the valence is also available but not included in our recommended tables.

\makefigure{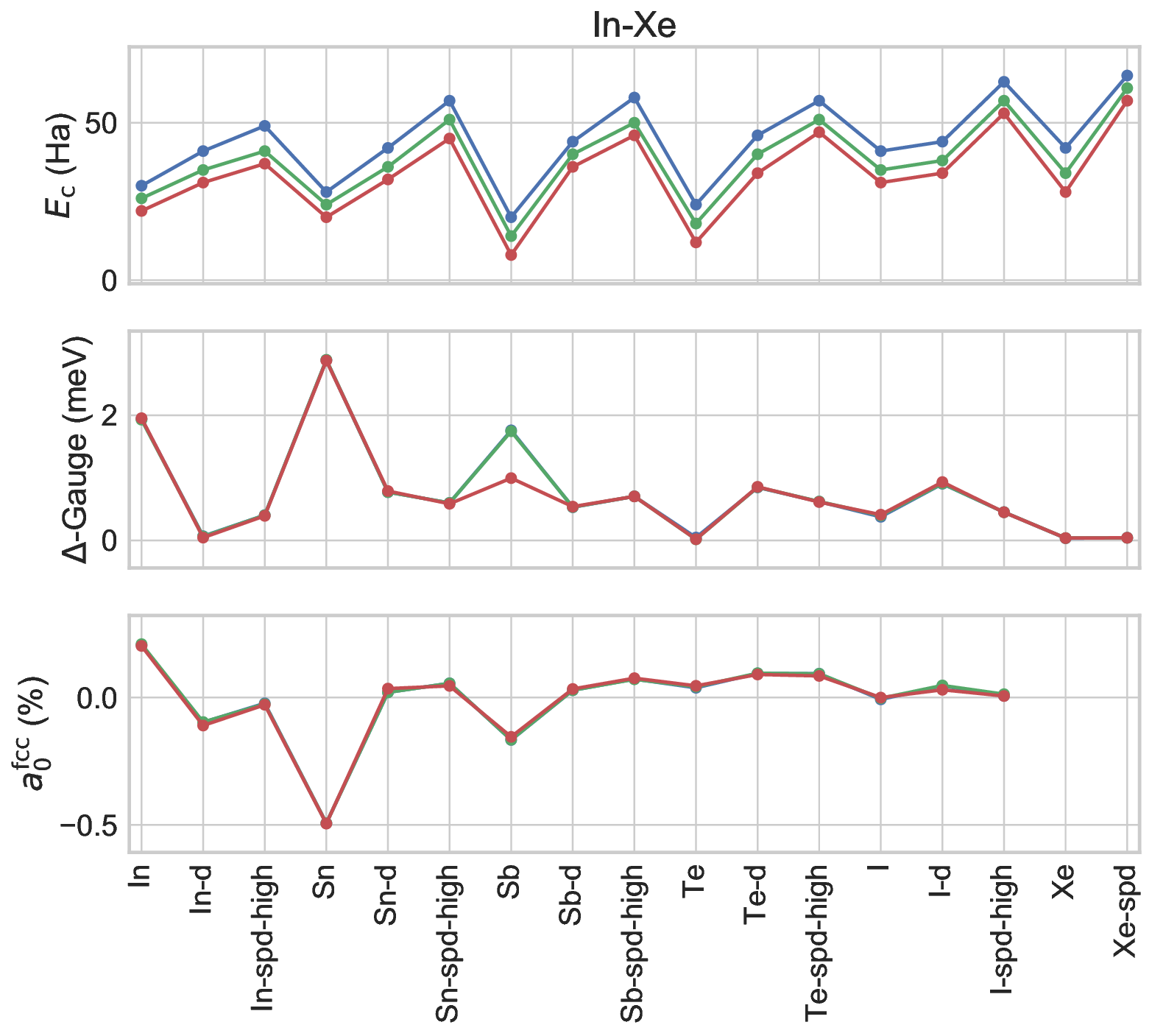}{The main test results for the In to Xe PSPs. The blue, green, and red data are calculated at the high, normal, and low $E_\mathrm{c}$ hints respectively.}

\subsection{Cs, Ba} 

The pseudopotentials for Cs and Ba both have the $5s$ and $5p$ states in the valence. For Cs, the transferability could be improved by adding explicit $f$ projectors. For Ba, also a PSP is provided based on a reference state in which a $6s$-electron is excited to the $5d$-state. This version improves the \df\: results but at the same time worsens the GBRV test results to a similar degree, see Fig.~\ref{Cs-Ba}. 

\makefigure{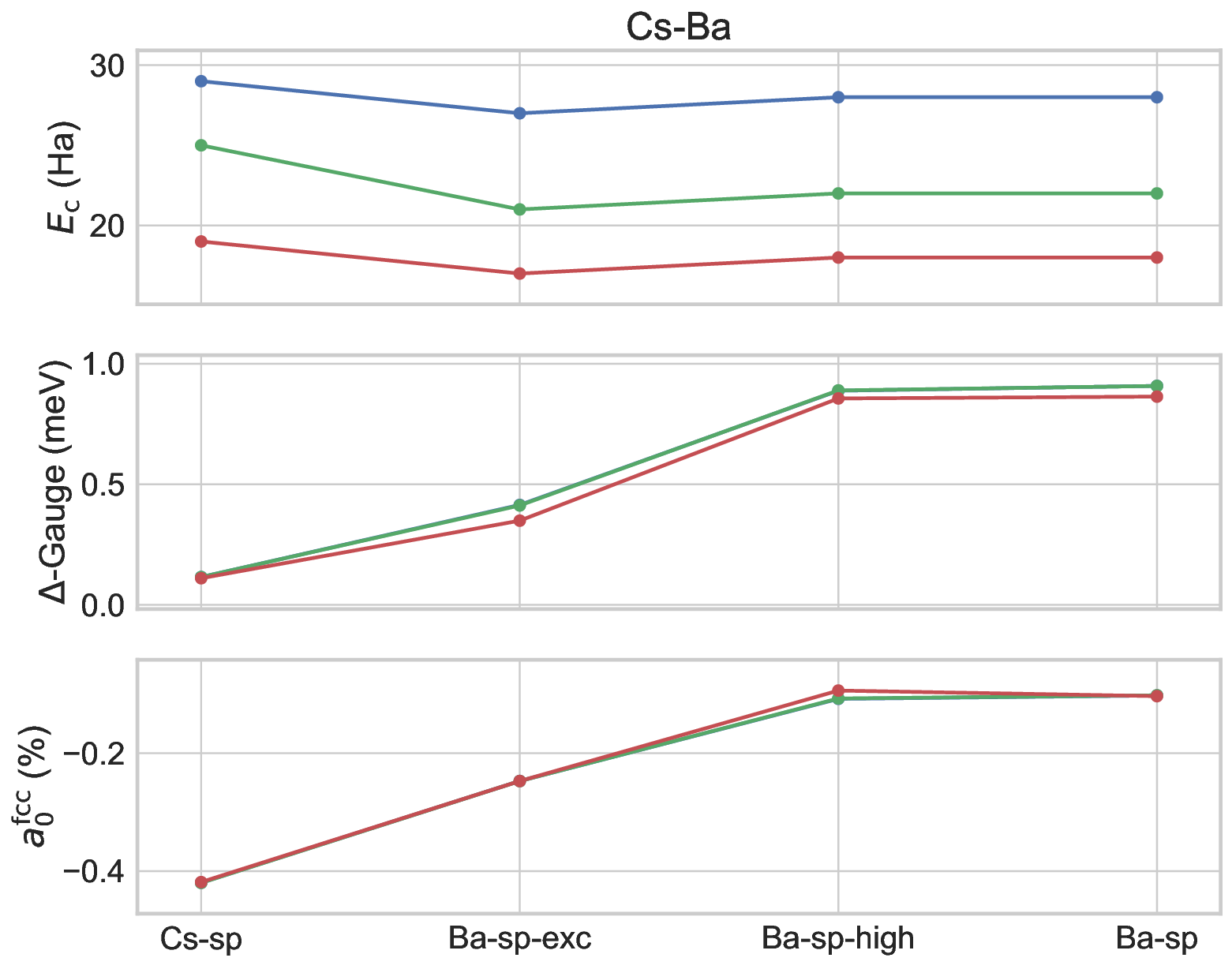}{The main test results for the Cs and Ba PSPs. The blue, green, and red data are calculated at the high, normal, and low $E_\mathrm{c}$ hints respectively.}

\subsection{5d transition metals}  

For the $5d$ transition metals we observed that including only the $5d$ in the valence led to PSPs that sometimes have good test results for \df\: and GBRV but tend to have ghost states only a few eV above the Fermi level. For this reason, we always include the $5s$ and $5p$ in the valence partition.

An additional difficulty in the series of the $5d$ transition metals is that in PBE the $4f$ states lie in the the same energy range as the $5s$ and $5p$ states. For Hf and Ta, the $4f$ even lie above the $5p$ states. Indeed for Hf the agreement with the AE reference for both the \df\: and GBRV tests improves significantly if, besides the $5s$ and $5p$ also the $4f$ is taken into the valence partition, see Fig.~\ref{5d-transition-metals}. For Ta, this still improves the \df\: results significantly but the GBRV results worsen. For W the changes are rather small. 

For all PSPs for the $5d$ transition metals it turned out to be beneficial to include explicit $f$-projectors even when the $4f$ electrons are frozen in the core.

Finally, we note that, although for W-Hg the ground state properties can be described well enough with the $4f$ frozen, for optical properties and $GW$ this may not be the case. This is the case even for elements like Au where the $4f$ electrons lie about 3 Ha below the Fermi level. 

\makefigure{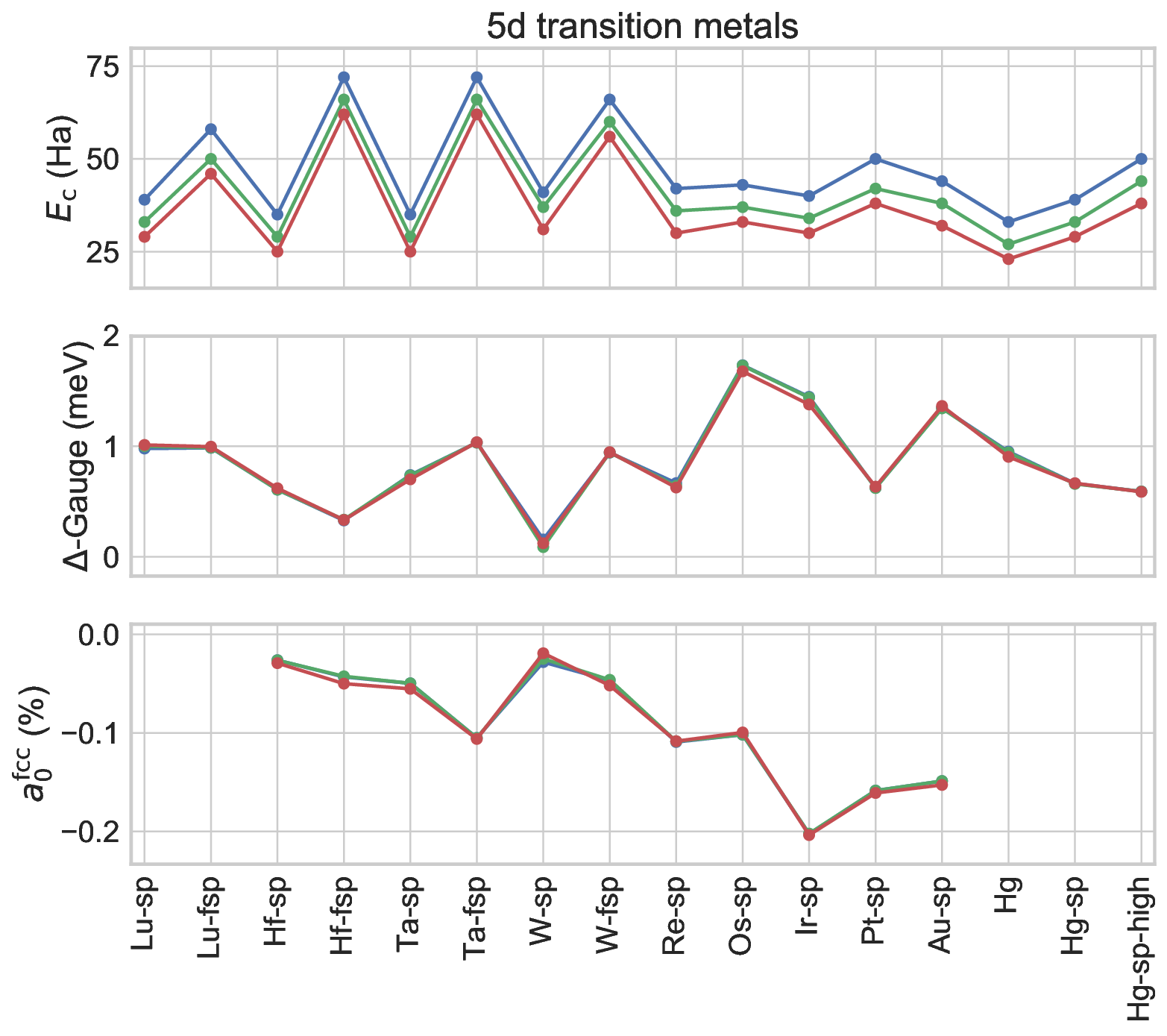}{The main test results for the $5d$ transition metal PSPs. GBRV reference data is not available for Lu and Hg. The blue, green, and red data are calculated at the high, normal, and low $E_\mathrm{c}$ hints respectively.}

\subsection{Lanthanides}

In crystalline systems the lanthanides usually occur in the 3+ oxidation state with three electrons donated to an anion. 
Typical examples of lanthanides with 3+ oxidation state are given by their nitrides.  
Since in standard KS theory, GGA or LDA, the strongly localized $f$ states are not described correctly, these 3+ pseudopotentials with $4f$ electrons frozen in the core offer a convenient solution.
They are all generated with the valence configuration: $5s^25p^65d^16s^2$.
It should be stressed, however, that these pseudopotentials are supposed to be used only if the $f$ electrons are not important in the physics of the crystal (e.g. magnetism, bonding, etc). These PSP will be mostly useful when only the steric effect of the lanthanide is of importance. 
Due to this limitation the lanthanide PSPs are not part of the predefined tables standard and stringent.
PSPs for lanthanides with $4f$ states in valence are currently being developed but testing these correctly is a topic on its own and will be presented elsewhere.

For La, GBRV reference results are available, our La-sp PSP underestimates the BCC lattice parameter by 0.1\%, no \df\: reference is available for La. 
For Lu the availability of reference results is opposite; we find a \df\: of 1.0 for our Lu-fsp, Lu is the only exception where the $4f$ is included in the valence. 
The hints we derive for these two elements based on the convergence of these tests are 50, 55, and 65 and 46, 50, and 58 Ha (low, normal, high) for La and Lu respectively. 

For the other lanthanides, no GBRV or \df\: reference data are available. 
The PSPs presented here are therefore tested by comparing the relaxed lattice parameters of their nitride rock salt structures with those obtained from PAW calculations. 
Figure~\ref{lantanide_lp} compares the lattice parameters obtained using our PSPs to those obtained using \textsc{VASP}~\cite{vasp1} with comparable 3+ PAW data sets.\footnote{In the \textsc{VASP} calculations we use a Gaussian smearing with 1~mH broadening, precision accurate settings, and converge the lattice parameters with respect to the kinetic energy cutoff.}

\makefigure{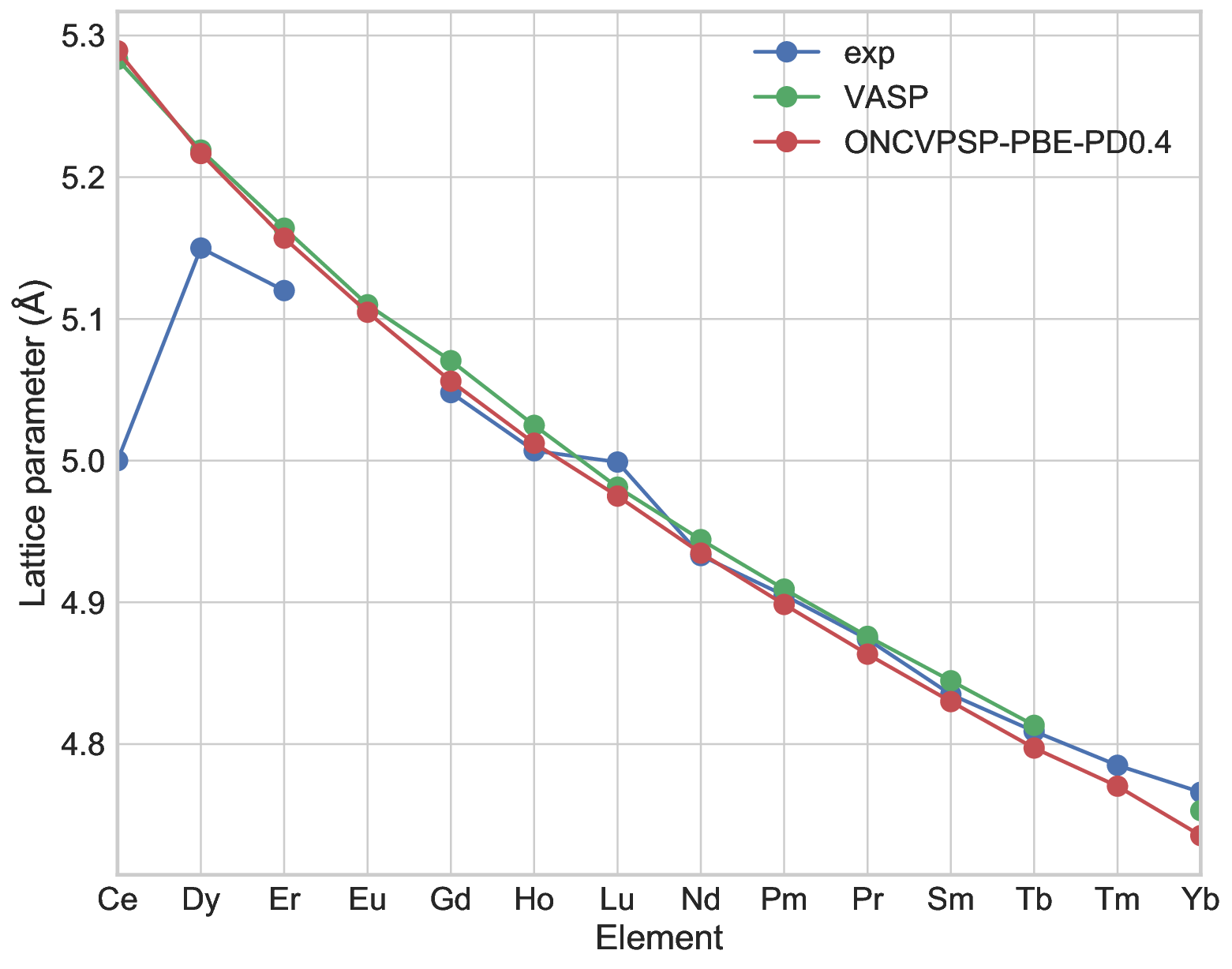}{Lattice parameters of the nitrides of the lanthanide series. Comparison of \ourtable\: to \textsc{VASP} results obtained with comparable PAW data sets. For reference also the experimental results are shown.}

In general we observe a very decent agreement between the \ourtable\: and the \textsc{VASP} results. 
Moreover, comparing to the experimental results we conclude that for the structural properties of rocksalt nitrides, the $4f$ states of Sm-Lu can indeed be frozen in the core.

\subsection{Tl, Pb, Po, At, Rn} 

For the final set of elements, Tl--Rn, the pseudization of $6p$ valence electrons is not very demanding. We provide versions with the $5d$ in the valence in the standard table and versions with $5s$ and $5p$ in the valence as well in the high table. Both for the \df\: and GBRV tests the results are good and also converge quickly, see Fig.~\ref{Tl-Rn}. 

\makefigure{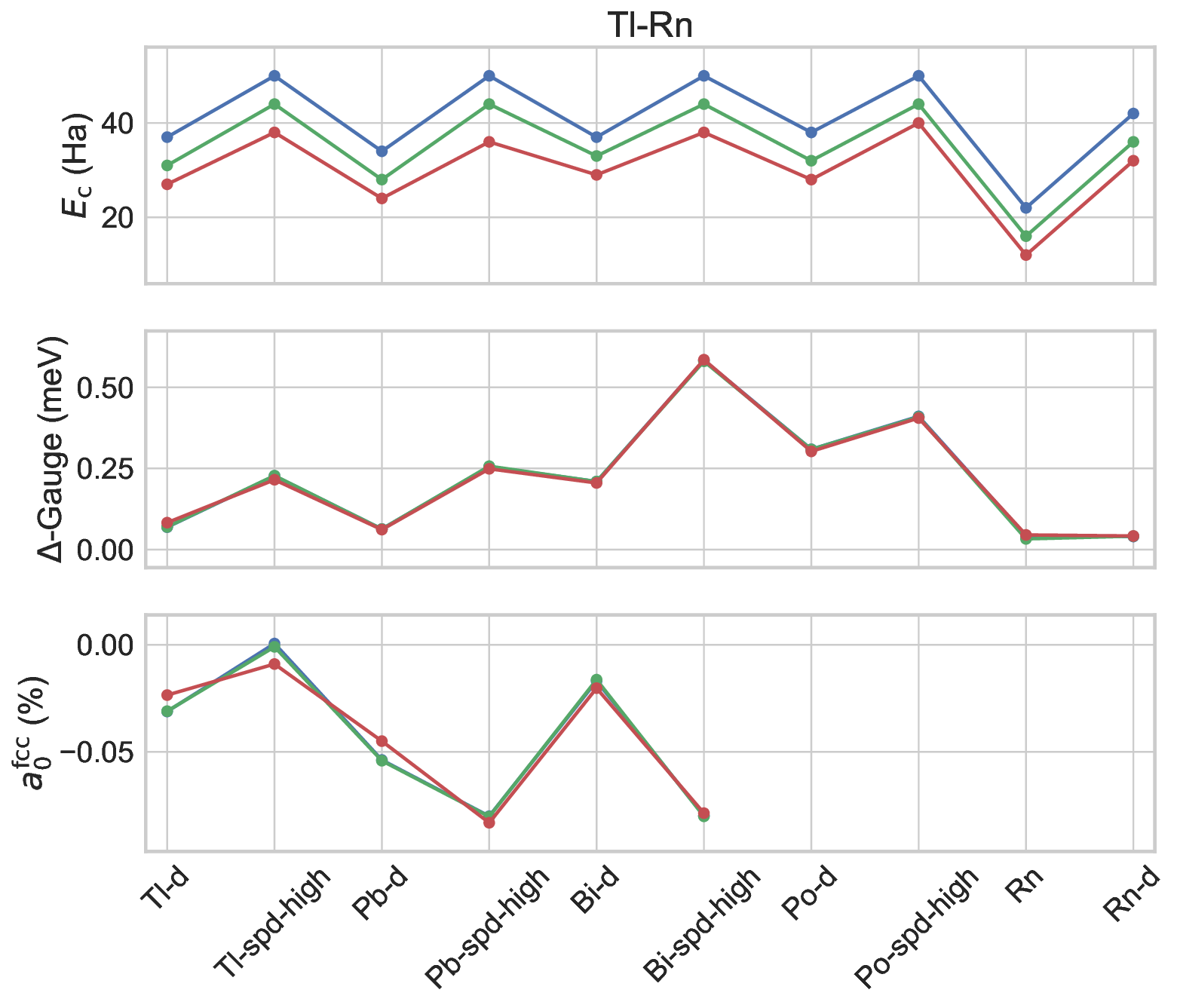}{The main test results for the Tl to Rn PSPs. GBRV reference data is not available for Po and Rn. The blue, green, and red data are calculated at the high, normal, and low $E_\mathrm{c}$ hints respectively.}

\section{Conclusions}

In this paper we have presented the \textsc{PseudoDojo} project, a framework for developing, testing and storing pseudopotentials, and discussed our \ourtable: an 84 element table of PBE norm-conserving pseudopotentials. The \textsc{PseudoDojo} is interfaced with \textsc{ONCVPSP}~\cite{PhysRevB.88.085117,PhysRevB.95.239906} to  generate the PSPs and \textsc{ABINIT}~\cite{Gonze2009,Gonze2016} via the \textsc{AbiPy} package for running the tests. The PSP files are available on the \textsc{PseudoDojo} web-interface at www.pseudo-dojo.org in the psp8, UPF2, and PSML 1.1 formats.

The \textsc{PseudoDojo} toolkit contains a graphical interface to the \textsc{ONCVPSP}~\cite{PhysRevB.88.085117,PhysRevB.95.239906} code. It enables the generation of (series of) pseudopotentials and the preparation of tests.

The validation part of the \textsc{PseudoDojo} consists of a series of 7 tests in crystalline environments: \df,~\cite{Lejaeghere_2013} \dfp,~\cite{Jollet_2014} GBRV-FCC, GBRV-BCC, GBRV-compound,~\cite{Garrity_2014} ghost state detection, and phonons at $\Gamma$, all executed using \textsc{ABINIT}.\cite{Gonze2009,Gonze2016} By studying the correlation between the results for the different tests we show that these form a complementary set. Only the GBRV-FCC and BCC show a strong correlation, such that performing both does not increase the amount of information. 

The present version of the \ourtable\: contains a total of 141 PSPs and defines two tables, with standard and stringent accuracy. Both tables contain one PSP per element. For the final set of PSPs a total of around 70.000 calculations have been performed during the testing process. All these calculations have been performed using the \textsc{PseudoDojo} tools building on the high-throughput framework of the \textsc{AbiPy} project. 

In the development of the \ourtable\:, valuable insights were obtained concerning the effects of the core-valence partitioning and the non-linear core corrections on the stability, convergence, and transferability of norm-conserving pseudopotentials. 

{\bf Non-linear core corrections} - PSP that have the $1s$ frozen in the core and the $2s$ and $2p$ completely filled (included in the valence partition) do not improve upon adding non-linear core corrections. Often they even become unstable (small changes in the unit cell volume or $E_\mathrm{c}$ lead to drastic changes in the total energy).
For the magnetic $3d$ transition metals adding well-balanced non-linear core corrections dramatically improves the results on magnetic systems. Non-magnetic systems are much less sensitive and also perform well without non-linear core corrections.
In some cases the model core charges for non-linear core correction can be quite localized. These hard models reproduce the AE results very well, and can have beneficial effects on ground state properties, but may render the PSP difficult to converge, especially in DFPT calculations.

{\bf Core-valence partitioning} - In the fifth row main group elements, the description of the unoccupied space improves clearly by {\em increasing} the valence partition. Including $4d$ alone leads to actual ghost states in the range of 20-80~eV above the Fermi level. Including the full $n=4$ shell removes all signs of ghost states up to several hundreds of eV. The exception is Xe for which putting the $n=4$ shell in the valence partition does not lead to any negative effect. A similar situation arises in the $5d$ transition metals. Including only the $5d$ in the valence partition for these elements leads to ghost states just above the Fermi level. Despite the good results obtained for the \df, these pseudopotentials are not transferable
and can perform poorly if used in other crystalline environments. 

{\bf Extra projectors} - In both the second row B-F and fifth row transition metals with frozen $4f$, the PSP are improved by the addition of additional projectors, $d$ and $f$, respectively.

Supplementary material: PD\_v0.4\_supplementary-data-and-tests: HTML version of the Jupyter Notebook performing the statistical analysis presented in this work, PD\_v0.4\_supplementary-correlations-elements: HTML version of the Jupyter Notebook performing the element wise comparison and correlation studies between the tests, PD\_v0.4\_supplementary-GBRV-compounds-standard: HTML version of the Jupyter Notebook performing statistical analysis of the GBRV compound tests and the lanthanide nitride lattice parameter comparison.

\section*{Acknowledgements}
Financial support was provided from FRS-FNRS through the PDR Grants T.1031.14 (HiT4FiT) for MJVS, T.0238.13 (AIXPHO) for MG, and T.1077.15 (Transport in novel vdW heterostructures) for MJV. The work was supported by the Communaut\'{e} Fran\c{c}aise de Belgique through the BATTAB project (ARC 14/19-057) and AIMED (ARC 15/19-09). We also thank the C\'ECI facilities funded by F.R.S-FNRS (Grant No. 2.5020.1) and Tier-1 supercomputer of the F\'ed\'eration Wallonie-Bruxelles funded by the Walloon Region (Grant No. 1117545). The authors thank J.M. Beuken and Y. Pouillon for technical support, B. Van Troeye, J.-J. Adjizian, A. Miglio and the other members of the NAPS group at the UCLouvain for feedback on the earlier versions of the pseudopotentials, M. Stankovski, J. Junquera, and A. Garc\'{i}a for useful discussions, and M. Ueshiba for inspiration.

\bibliography{bibfile.bib}

\end{document}